\documentclass[pra,showpacs,preprintnumbers,amsmath,amssymb,twocolumn]{revtex4}
\usepackage{graphicx}% Include figure files  graphicx.sty
\usepackage{dcolumn}% Align table columns on decimal point
\usepackage{bm}% bold math
\usepackage{color}
\usepackage{layout}
\usepackage{verbatim}

\newcommand{\ket}[1]{\left|#1\right>}
\newcommand{\bra}[1]{\left< #1 \right|}
\newcommand{\beq}{\begin{equation}}
\newcommand{\eeq}{\end{equation}}
\newcommand{\bep}{\begin{pmatrix}}
\newcommand{\eep}{\end{pmatrix}}

\newcommand{\id}{\mathbb{I}}

\newcommand{\Vr}{V_{r}}
\newcommand{\Vm}{V_{m}}
\newcommand{\Vl}{V_{l}}
\newcommand{\JL}{J_{l}}
\newcommand{\JR}{J_{r}}

\newcommand{\h}[1]{\mathcal{#1}}
\newcommand{\dBLM}{\Delta B_{l}}
\newcommand{\dBMR}{\Delta B_{r}}

\newcommand{\vrf}{v_{\rm{rf}}}

\newcommand{\UDU}{\ket{\uparrow\downarrow\uparrow}}
\newcommand{\DUU}{\ket{\downarrow\uparrow\uparrow}}
\newcommand{\UUD}{\ket{\uparrow\uparrow\downarrow}}
\newcommand{\UUU}{\ket{\uparrow\uparrow\uparrow}}
\newcommand{\SL}{\ket{S_l}}
\newcommand{\SR}{\ket{S_r}}
\newcommand{\SM}{\ket{1}}
\newcommand{\TL}{\ket{T_l}}
\newcommand{\TR}{\ket{T_r}}
\newcommand{\TM}{\ket{0}}
\newcommand{\Q}{\ket{Q}}
\newcommand{\Qp}{\ket{Q_{+}}}
\newcommand{\ES}[1]{E_{#1}}
\newcommand{\ET}[1]{(\id - \ES{#1})}

\newcommand{\eps}[2]{\varepsilon_{#2}^{\rm{#1}}} 
\newcommand{\tauJM}[1]{\tau_{\rm{#1}}} 
\newcommand{\rhoL}{\ket{S_l}\!\!\bra{S_l}}
\newcommand{\rhoR}{\ket{S_r}\!\!\bra{S_r}}
\newcommand{\HJ}[1]{H_{J}(\eps{#1}{})}
\newcommand{\Heff}[2]{\mathcal{H}(\eps{#1}{#2})}
\newcommand{\sigL}{\JL}
\newcommand{\sigR}{\JR}
\newcommand{\rhoZZ}{\ket{0}\!\!\bra{0}}
\newcommand{\rhoOO}{\ket{1}\!\!\bra{1}}
\newcommand{\rhoZO}{\ket{0}\!\!\bra{1}}
\newcommand{\rhoOZ}{\ket{1}\!\!\bra{0}}
\newcommand{\rhoQQ}{\ket{Q}\!\!\bra{Q}}

\newcommand{\imRhoZO}{\mathrm{Im}[\rhoZO]}
\newcommand{\reRhoZO}{\mathrm{Re}[\rhoZO]}
\newcommand{\rhoThree}{\hat{R}_{r}\!\ket{S_l}\!\!\bra{S_l}\!\hat{R}_{r}^\dag}
\newcommand{\rhoFour}{\hat{R}_{l}\!\ket{S_r}\!\!\bra{S_r}\!\hat{R}_{l}^\dag}
\begin{document}

\title{Self-Consistent Measurement and State Tomography of an Exchange-Only Spin Qubit}

\date{\today}

\author{J.~Medford$^{1}$}
\author{J.~Beil$^{1,2}$}
\author{J.~M.~Taylor$^{3}$}
\author{S.~D.~Bartlett$^{4}$}
\author{A.~C.~Doherty$^{4}$}
\author{E.~I.~Rashba$^{1}$}
\author{D.~P.~DiVincenzo$^{5,6}$}
\author{H.~Lu$^7$}
\author{A.~C.~Gossard$^7$}
\author{C.~M.~Marcus$^{1,2}$}
\affiliation{$^1$Department of Physics, Harvard University, Cambridge, Massachusetts 02138, USA\\
$^2$Center for Quantum Devices, Niels Bohr Institute, University of Copenhagen, Universitetsparken 5, DK-2100 Copenhagen, Denmark\\
$^3$Joint Quantum Institute/NIST, College Park, MD, USA\\
$^4$Centre for Engineered Quantum Systems, School of Physics, The University of Sydney, Sydney, NSW 2006, Australia\\
$^5$Institute for Quantum Information, RWTH Aachen University, 52056 Aachen, Germany\\ 
$^6$Dept. Theoretical Nanoelectronics, PGI, Forschungszentrum Juelich, 52425 Juelich, Germany\\
$^7$Materials Department, University of California, Santa Barbara, California 93106, USA
}

\begin{abstract}
We report initialization, complete electrical control, and single-shot readout of an exchange-only spin qubit. Full control via the exchange interaction is fast, yielding a demonstrated 75 qubit rotations in under 2~ns. Measurement and state tomography are performed using a maximum-likelihood estimator method, allowing decoherence, leakage out of the qubit state space, and measurement fidelity to be quantified. The methods developed here are generally applicable to systems with state leakage, noisy measurements, and non-orthogonal control axes. \end{abstract}

%\pacs{Valid PACS appear here}

\maketitle

%\textit{Introduction}--- 
\textbf{Nanoelectronics show great promise as a quantum information platform, in particular as superconducting qubits~\cite{nakamura1999coherent,Chiorescu21032003,PhysRevLett.89.117901,PhysRevA.76.042319, 2012arXiv1211.0322} and spin qubits in semiconductors~\cite{PettaScience,koppens2006driven,nowakSO,gaudreau2011coherent}. One or two electron spin qubits use, respectively, oscillating magnetic~\cite{koppens2006driven} or electric fields \cite{nowakSO,nadj2010spin},  or quasi-static Zeeman field gradients~\cite{LairdESR,  pioro2008electrically,folettiTomography, pettaZener}, to achieve full qubit control. Adding a third spin provides exchange-driven qubit rotations along two axes, hence full control of spin information via electrostatic gating only~\cite{divincenzo2000universal,LairdTdot,gaudreau2011coherent,hsieh2012physics,2012arXiv1211.0417M,west2012exchange}.}

\textbf{The three-electron exchange-only spin qubit has a more complicated level structure than its one- and two-electron counterparts~\cite{LairdTdot, hsieh2012physics, 2012arXiv1211.0417M}, providing, for example, multiple initialization states, but also allowing leakage out of the qubit state space. Here, we characterize the performance of the three-electron spin qubit by performing measurement and state tomography~\cite{lundeen2008tomography,brida2012quantum, 2012arXiv1211.0322}. Measurement tomography allows accurate state tomography in the presence of noisy measurements and leakage.} 

\begin{figure}%[hb!]
\includegraphics[width = 3 in]{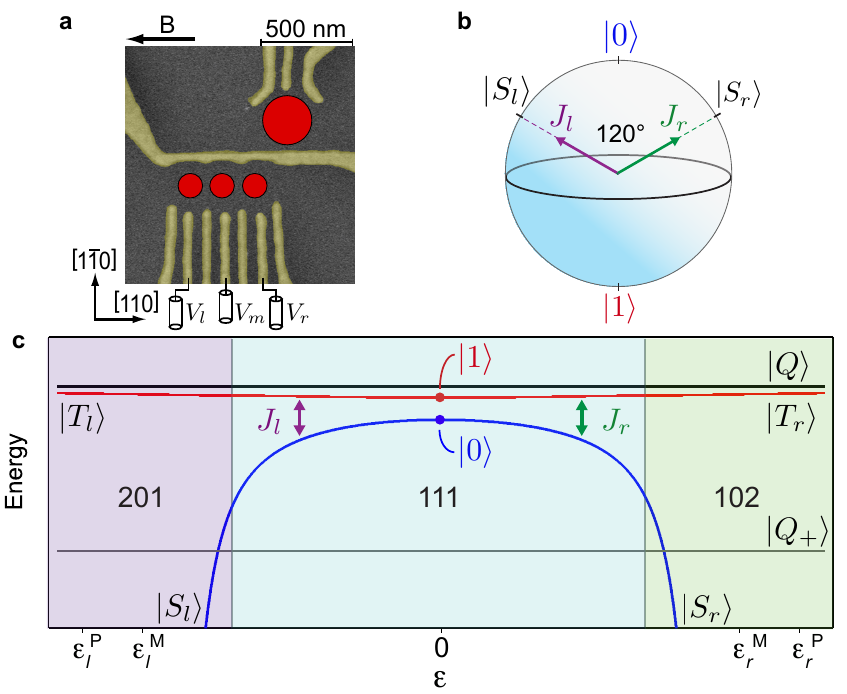}
\caption[Device, qubit Bloch sphere and spectrum]{\label{FigDevice}{\bf Device, qubit Bloch sphere and spectrum.}~(a)~False color micrograph of lithographically identical device with locations of triple dot (smaller red circles) and sensor dot (larger red circle). Gate voltages $V_{l}$ and $V_{r}$ set the charge occupancy of left and right dot as well as the detuning, $\eps{}{}$ of the qubit. (b) A Bloch sphere representation of the qubit with control axes $\JL$ and $\JR$ indicated, as well as two initialization states, $\SL$ and $\SR$.  (c) Energy levels as a function of detuning for the lowest energy states~\cite{LairdTdot}. The red and blue levels form the logical subspace inside 111, with the logical states $\TM$ and $\SM$ denoted at the detuning at which they are the eigenstates of the system. Each state has a spin-split partner state with opposite spin projection, not shown. Values of $\epsilon$ for preparation (P) and measurement (M) in 201 and 102 are indicated.}
\end{figure}

\begin{figure}[t]
\includegraphics{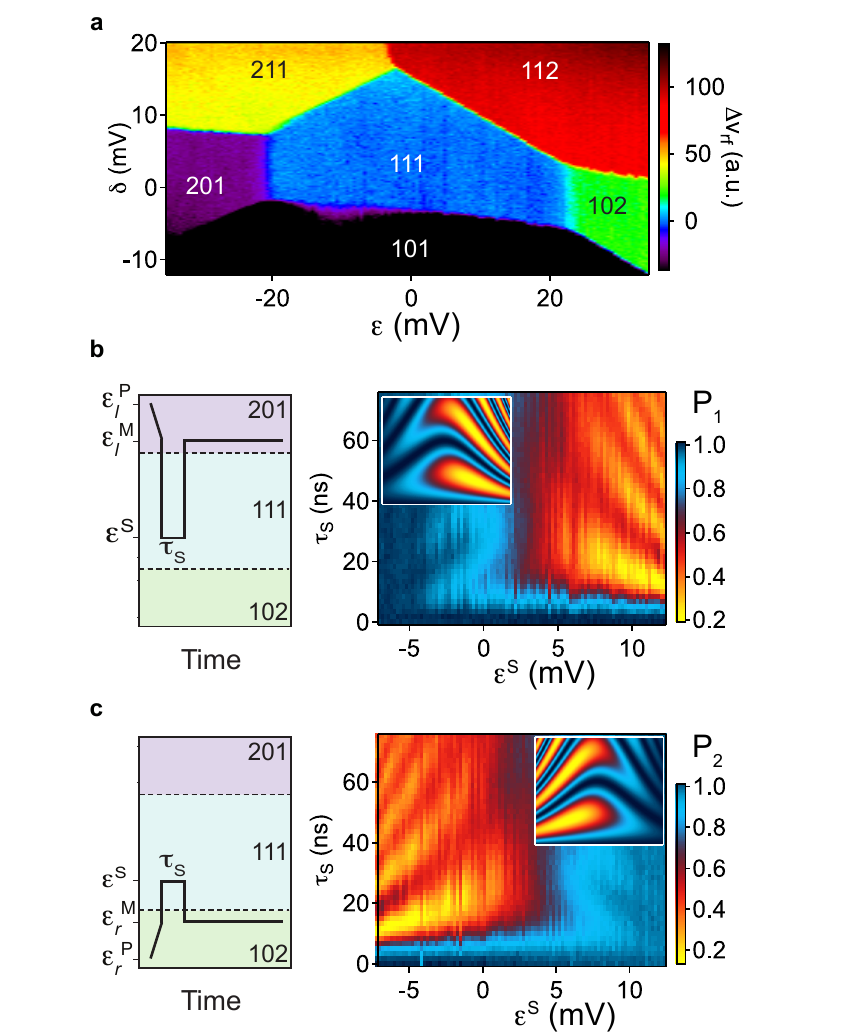} %[width = 3.2 in]
\caption[Charge stability diagram and rotations around two axes]{
\label{FigJeps}{\bf Charge stability diagram and rotations around two axes.}
~(a)~Triple dot charge occupancy N$_{l}$ N$_{m}$ N$_{r}$ as a function of $\Vl$ and $\Vr$ in and near the 111 regime; $\eps{}{} = (\Vr -\Vr^{0})/2 - (\Vl-\Vl^0)/2$, $\delta =  (\Vr -\Vr^{0})/2 + (\Vl-\Vl^0)/2$~~\cite{detuningFootnote}. 
The charge occupancy is measured using the change in the reflected rf signal, $\Delta v_{\rm{rf}}$, incident on the proximal sensor. 
(b)~Schematic of a pulse sequence that prepares $\SL$ in 201, and transfers that state to 111, by moving along $\eps{}{}$ at $\delta=0$. The sequence then waits at $\eps{S}{}$ for a time $\tauJM{S}$, and returns to 201 for measurement.  The probability, $P_1$, of remaining in the initial state, $\SL$, is plotted as a function of pulse position and wait time. Positive $\eps{}{}$ brings the state closer to 102, while negative $\eps{}{}$ brings the state closer to 201.
(c) Schematic of a pulse sequence, along with a plot of the probability of remaining in $\SR$ for an excursion to the separation point $ \eps{S}{}$ for a time $\tauJM{S}$. (insets) Model of qubit evolution as a function of exchange. No noise has been included.}
\end{figure}

A three-electron linear triple quantum dot was formed by Ti/Au electrostatic gates patterned on a GaAs/AlGaAs heterostructure with the two-dimensional electron gas 110 nm below the surface (see Fig.~\ref{FigDevice}(a)). Left and right plunger voltages, $\Vl$ and $\Vr$, controlled electron occupation of each dot. All manipulations kept a three electron total, with the arrangement, $(N_l\, N_m\, N_r)$, set by the detuning parameter, $\eps{}{} = (\Vr -\Vr^{0})/2 - (\Vl-\Vl^0)/2$, where $\eps{}{} = 0$ is defined as the center of 111 (see Fig.~\ref{FigJeps}(a)). 

Three electrons have eight possible spin states, four with total spin $S = 3/2$, and four with $S=1/2$~\cite{Buchachenko,LairdTdot}. An external magnetic field splits the eight states into four subspaces with spin projection, $m_S = \pm3/2, \pm1/2$. The linear geometry allows two exchange interactions, which lower the energy of singlet-like pairs within the $S=1/2$, $m_S = \pm1/2$ subspaces. In particular, tunneling between the left and middle dots opens a splitting, $\JL(\eps{}{})$, between the left singlet-like~\cite{singletLikeFootnote} state $\SL = \frac{1}{\sqrt{2}}(\UDU-\DUU)$ and the left triplet-like state $\TL = \frac{1}{\sqrt{6}}(\DUU+\UDU-2\UUD)$. $\JL(\eps{}{})$ increases as the detuning is shifted towards the 201 charge state. Tunneling between right and middle dots similarly opens a splitting $\JR(\eps{}{})$ between $\SR = \frac{1}{\sqrt{2}} (\UUD-\UDU)$ and $\TR =  \frac{1}{\sqrt{6}}(\UUD+\UDU-2\DUU)$ which increases as $\eps{}{}$ is shifted towards 102. 

The logical qubit space is chosen to be in the $S=1/2$, $m_S = +1/2$ subspace~\cite{ZeemanFootnote}, where gate voltages control the energy spectrum.  The logical qubit states, $\TM =  \frac{1}{\sqrt{6}}\left(\UUD + \DUU - 2\UDU  \right)$ and $\SM =  \frac{1}{\sqrt{2}}\left(\UUD - \DUU\right)$, are eigenstates in the center of 111, with $\JL(\eps{}{})=\JR(\eps{}{})$.  Two states with $S=3/2$ couple into the logical subspace through Zeeman field gradients. Longitudinal gradients couple the qubit space to the $S=3/2, m_S=1/2$ state, $\Q = \frac{1}{\sqrt{3}}\left( \UUD + \UDU + \DUU \right)$. The state $\Q$ is a spin symmetric state, being triplet-like for both left-middle and middle-right pairs of spins, and is the dominant leakage state for this qubit.  Leakage into the $S=3/2, m_S=3/2$ state, $\Qp = \UUU$, is suppressed by a large Zeeman field except at two anticrossings. By traversing these anticrossings diabatically---unlike in previously triple-dot experiments~\cite{LairdTdot,gaudreau2011coherent}--- leakage into $\Qp$ can be made negligible.

The left singlet-like state, $\SL$, is prepared by moving to $\eps{P}{l}$ in 201, and briefly moving near the 201-101 charge transition border to promote rapid relaxation to the ground state. The right singlet-like state, $\SR$, is similarly prepared by moving to $\eps{P}{r}$ in 102 and pulsing near the charge border. The excursions to these charge borders during initialization are the only departures from $\delta =  0$, where $\delta = (\Vr -\Vr^{0})/2 + (\Vl-\Vl^0)/2$ defines the center line of 111 (see Fig.~\ref{FigJeps}(a)).

Arbitrary qubit states are determined by projection onto $\SL$ or $\SR$. Projection onto $\SL$ is accomplished by moving to the left measurement point, $\eps{M}{l}$, where $\SL$ can move to 201 while $\TL$ and $\Q$ remain trapped in  111~\cite{LairdTdot}. 
During the measurement, an rf excitation is applied across the sensor quantum dot. The reflected signal is demodulated using homodyne detection~\cite{Reilly_APL07} and integrated for $\tauJM{M} = 50~\mu$s, resulting in a signal corresponding to either the 201 or 111 charge state. Projection of the qubit state onto $\SR$ is carried out in a similar way at measurement point $\eps{M}{r}$ in 102.

Effects of exchange interactions, $\JL(\eps{}{})$ and $\JR(\eps{}{})$, on qubit dynamics are modeled by an effective Hamiltonian 
\beq
\HJ{} = \JL(\eps{}{})\sigma_l+ \JR(\eps{}{})\sigma_r,
\eeq
where $\sigma_l \equiv(\sqrt{3}\sigma_x-\sigma_z)/4$, $\sigma_r \equiv(-\sqrt{3}\sigma_x-\sigma_z)/4$, and $\sigma_x$ and $\sigma_z$ are Pauli matrices in the logical basis $\{\TM$, $\SM\}$. As illustrated in Fig.~\ref{FigDevice}(b), $\JL(\eps{}{})$ and $\JR(\eps{}{})$ drive rotations about axes that are 120$^{\circ}$ apart on the Bloch sphere. 
In what follows, we use the terms $\JL$ and $\JR$ rotations and axes in the spirit of this model.

To demonstrate two-axis control and readout, as well as to test the applicability of the simple model, Eq.~(1), we first initialize the system in the $\SL$ state and separate the electrons into 111 at a detuning $\eps{S}{}$ for a time $\tauJM{S}$, where the qubit evolves under $\HJ{S}$ [Fig.~\ref{FigJeps}(b)]. The qubit is then pulsed to $\eps{M}{l}$ to measure the projection of the evolved state onto $\SL$, which we determine by measuring the singlet return probability on the left, $P_1$, over an ensemble of repeated experiments. Pulsing instead to $\eps{M}{r}$ allows for a measurement of the projection onto $\SR$, which when averaged over an ensemble gives the singlet return probability on the right, $P_2$.

Figure \ref{FigJeps}(b) shows that for states initialized in $\SL$, there is a rapid oscillation of the measured $P_1$ as a function of $\tauJM{S}$ at positive detunings $\eps{S}{}$, $\JL(\eps{S}{})\ll\JR(\eps{S}{})$, and a roughly constant $P_1\sim 1$ at negative detunings, $\JL(\eps{S}{})\gg \JR(\eps{S}{})$. The reverse is true for states prepared as $\SR$ in Fig.~\ref{FigJeps}(c): $P_2\sim1$ at positive detunings while $P_2$ exhibits rapid oscillations as a function of $\tauJM{S}$ at negative detunings.

The insets of Fig.~\ref{FigJeps} show model calculations of $P_1$ = $|\bra{S_l}e^{-i\HJ{S}\tauJM{S}/\hbar}\ket{S_l}|^2$ and $P_2$ = $|\bra{S_r}e^{-i\HJ{S}\tauJM{S}/\hbar}\ket{S_r}|^2$, which agree well with experiment. These calculations neglect noise in $\JL(\eps{S}{})$ and $\JR(\eps{S}{})$ as well as fluctuations in local hyperfine fields. These contributions are considered in detail below. 

\begin{figure}%[t]
%\centering
\includegraphics{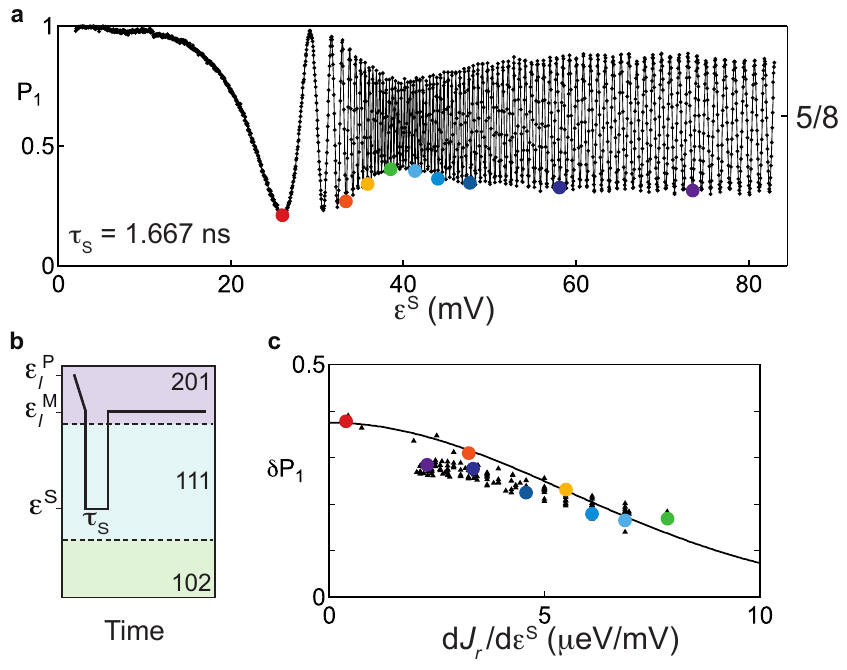}%[width = 2.5 in]
\caption[Fast rotation and visibility model]{
\label{FigFIDex}{\bf Fast rotation and visibility model.}~(a)~ Qubit precession as a function of detuning at the separation point, $\eps{S}{}$ for fixed pulse duration, $\tauJM{S} = 1.667$ ns. This pulse width is less than the rise time of both the coax and the pulse generator, leading to a significant pulse attenuation at the sample. (b) One cycle of the pulse sequence schematic used to generate rotation about the $\sigR$ axis. (c) Amplitude of oscillation, $\delta P_1$, as a function of $\mathrm{d}\JR/\mathrm{d}\eps{S}{}$, measured from the period of oscillations in (a). Theoretical form $\delta P_1 = \frac{3}{8} \,e^{-\alpha^2 (\mathrm{d}\JR/\mathrm{d}\eps{S}{})^2}$, where $\alpha = \tauJM{S}\,\sigma_{\eps{}{}}/\sqrt{2}$ \cite{Cywinski_PRB08} with single parameter, $\sigma_{\eps{}{}} = 450 ~\mu$V, fit over the first 29 oscillations, to the left of the green point ($4^\mathrm{th}$ from the left) dot in (a).}\end{figure}

%\textit{Exchange Noise}--- 
An exchange pulse can generate rapid qubit evolution on nanosecond time scales, faster than dynamics induced by other sources such as spin-orbit or hyperfine coupling. The short-pulse regime thus allows exchange and its noise to be examined in isolation from other sources of qubit dynamics.  Figure 3(a) shows $P_{1}$ for a short exchange pulse, $\tauJM{S} = 1.667$ ns, as a function of pulse amplitude, $\JR(\eps{S}{})$, over a range of phase $\phi =\tauJM{S}\JR(\eps{S}{})/\hbar$ from 0 to $\sim 158 \pi$, corresponding to a 47.4 GHz rotation.

At large positive $\eps{S}{}$, where $\JR(\eps{S}{})\gg\JL(\eps{S}{})$, the noiseless model predicts $P_1 = 5/8 + 3/8 \cos(\tauJM{S}\JR(\eps{S}{})/\hbar)$ for initial state $\SL$~\cite{AdiabatFootnote}. Experimental data agrees well with the 5/8 average [see Fig.~3(a)], but the observed oscillation amplitude, $\delta P_1$, is notably less than 3/8, with a distinct dip where phase varies most rapidly with $\eps{S}{}$, i.e., where $\mathrm{d}\JR/\mathrm{d}\eps{S}{}$ is largest. The reduced amplitude can be understood quantitatively as the result of averaging over exchange noise arising from noise in $\eps{}{}$, yielding $\delta P_1 = \frac{3}{8} \,e^{-\alpha^2 (\mathrm{d}\JR/\mathrm{d}\eps{S}{})^2}$, where $\alpha = \tauJM{S}\,\sigma_{\eps{}{}}/\sqrt{2}$ \cite{Cywinski_PRB08}. The period of oscillations in Fig.~3(a) gives a direct measurement of $\mathrm{d}\JR/\mathrm{d}\eps{S}{}$, leaving a single fit parameter, $\sigma_{\eps{}{}}$, the effective standard deviation of noise in $\eps{}{}$. Experiment and theory are in excellent agreement [Fig.~\ref{FigFIDex}(c)]. The fit value, $\sigma_{\eps{}{}} = 450 ~\mu$V, is only nominal, as it includes effects of finite coax bandwidth, making it larger than the actual $\eps{}{}$ noise in the system.

\begin{figure}%[t]
%\centering
\includegraphics{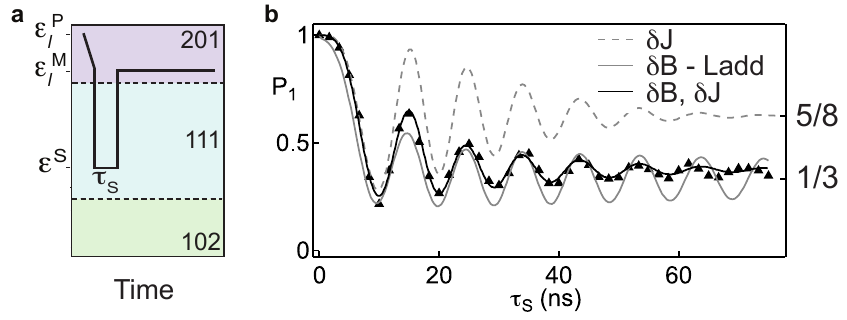}%[width = 2.5 in]
\caption[Effects of electrical and nuclear noise]{
\label{FigFIDnuc}{\bf Effects of electrical and nuclear noise.}~(a)  A schematic of the pulses for a rotation about the $\JR$ axis. (b)~Free induction decay (FID) as a function of $\tauJM{S}$. The dashed gray curve is a theory model of a rotation around the $\JR$ axis ($\JL=0$) in the presence of dephasing, the solid gray curve is a model of Ref.~\cite{Ladd12}, that accounts for the same rotation in the presence of dephasing and leakage due solely to nuclear magnetic field gradients. The solid black curve is fit to a numerical integration of the qubit evolution for a rotation around $\JR$, in the presence of quasistatic Gaussian distributions of nuclear gradients  $\dBLM$ and $\dBMR$, as well as quasistatic noise in $\JR$. $\JL$ was also allowed to be nonzero in the fit. Noise is evaluated as quasistatic, rather than incorporating unknown spectral densities. Here, $\JL = 20 \pm 8$ neV, $\JR = 388 \pm 2$ neV, and standard deviations $\sigma_B = 2.0 \pm 0.1$ mT, $\sigma_J = 19 \pm 2$ neV, with a turn-on of the exchange modeled as an exponential with time constant $\tauJM{} = 12.6 \pm 0.2$ ns.}
\end{figure}

In Fig~\ref{FigFIDnuc}, the free induction decay measured at longer separation times, $\tauJM{S}$, and at fixed detuning, $\eps{S}{}$, reveals the combined effects of exchange noise, which causes dephasing, discussed above, and Zeeman field gradients, which cause both dephasing and leakage out of the qubit space.  

Quasi-static longitudinal (effective) field differences between dots, $\dBLM = B^z_l - B^z_m$ and  $\dBMR = B^z_m - B^z_r$,  drive coherent evolution between $\TM$ and $\SM$. Gradients due to hyperfine fields appear static on the time scale of a single sequence of measurements, but execute a thermal random walk over an ensemble of measurements.  In addition,  Zeeman differences $g\mu_{\mathrm{B}}\dBLM$ and $g\mu_{\mathrm{B}}\dBMR$ comparable in magnitude to $\JL(\eps{}{})$ or $\JR(\eps{}{})$ will drive evolution into $\Q$, the leakage state. Here $\mu_{\mathrm{B}}$ is the Bohr magneton and $g \sim-0.4$ is the electron g factor.  Averaging over the entire nuclear ensemble during repeated measurements results in a damped oscillation towards the triplet outcome as the qubit dephases and leaks into $\Q$. By examining in detail the $\tauJM{S}$ dependence of $P_1$ at a fixed $\eps{S}{}$ in Fig.~\ref{FigFIDnuc}, and comparing it with theoretical models for low frequency exchange and Overhauser~\cite{Ladd12} noise, we conclude that nuclear fluctuations are the predominant source of noise in this system, with a standard deviation of 2.0 mT.

%\textit{Mini Echo}--- 
Low-frequency hyperfine noise~\cite{ReillySpectrum, koppensPhaseShift} can be compensated using dynamical decoupling \cite{BluhmT2, MedfordT2}. Unlike the situation in double quantum dots, however, a single-pulse echo cannot undo the effects of two hyperfine field gradients in the three-dot system\cite{west2012exchange}. Nevertheless, a single $\pi$-pulse can undo a \textit{portion} of the dephasing due both to nuclei and low-frequency exchange noise.

\begin{figure}%[t]
%\centering
\includegraphics{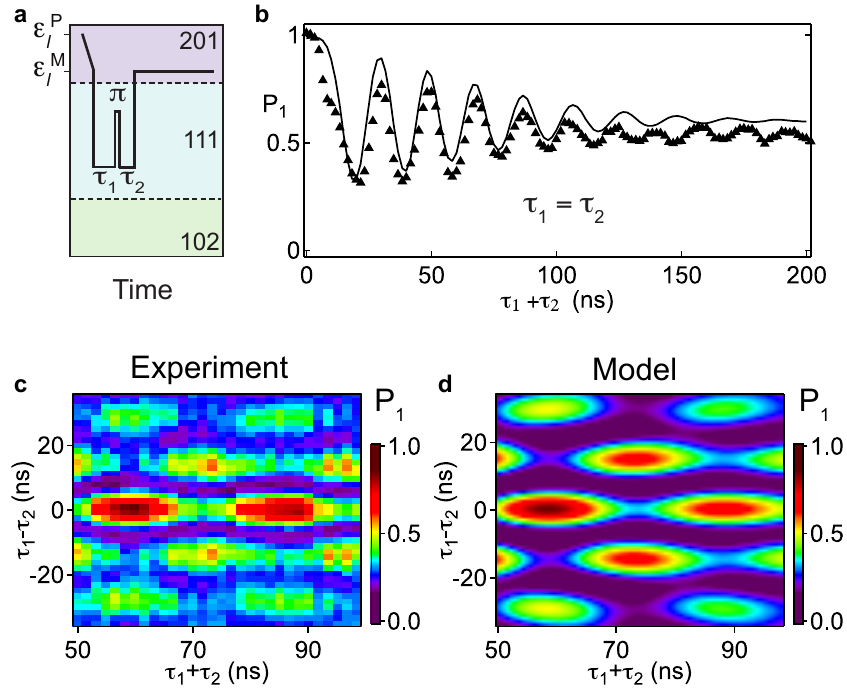}%[width = 2.5 in]
\caption[Dynamical decoupling]{
\label{FigMiniEcho}{\bf Dynamical decoupling.}~(a) A schematic for a three pulse sequence demonstrating an echo, where a prepared $\SL$ precesses around $\sigR$ for a time $\tauJM{1}$, performs a $\pi$-pulse around $\sigL$, and then precesses again around $\sigR$ for a time $\tauJM{2}$ before being measured in 201. (b) Probability $P_{1}$ of measuring $\SL$ for the pulse sequence depicted in (a), for the case $\tauJM{1}=\tauJM{2}$. The solid black curve is a model with noise from $\JR$, $\dBLM$, and $\dBMR$ during the dephasing times $\tauJM{1}$ and $\tauJM{2}$, and noise due to $\JL$, $\dBLM$, and $\dBMR$ during the $\pi$-pulse using parameters extracted from FID data in Fig.~\ref{FigFIDnuc}(b). (c) The results of a three pulse echo sequence, illustrated in (a), that maps the probability of remaining $\SL$ as a function of the total dephasing time ($\tauJM{1}+\tauJM{2}$), and the difference in time between the first free induction decay, $\tauJM{1}$, and the second, $\tauJM{2}$. (d) A model plot for (c) that averages over thermal distributions of nuclear gradients. See Supplementary Information for details of the calculation. }
\end{figure}

Single-pulse partial echo is demonstrated by preparing $\SL$ in 201, separating to 111 where the state rotates for a time $\tauJM{1}$ around $\JR$, followed by a $\pi$-pulse around $\JL$, followed by further rotation around $\JR$ for a time $\tauJM{2}$. The sequence is illustrated in Fig.~\ref{FigMiniEcho}(a).
$P_1$ shows robust oscillations as a function of both sum and difference of the dephasing times $\tauJM{1}$ and $\tauJM{2}$, similar to a Ramsey measurement with a refocusing pulse in the middle. The decay envelope in $\tauJM{1}+\tauJM{2}$ at $\tauJM{1}=\tauJM{2}$ in Fig.~\ref{FigMiniEcho}(b) gives a lower bound on the coherence time, $T_2 \sim 100$ ns, while the decay envelope in $\tauJM{1}-\tauJM{2}$ shown in Fig.~\ref{FigMiniEcho}(c) gives a dephasing time,  $T_2^*\sim25$ ns. The dephasing time in $\tauJM{1}-\tauJM{2}$ is consistent with FID times [Fig.~\ref{FigFIDnuc}(b)], while decay as a function of $\tauJM{1}+\tauJM{2}$ is extended by a factor of $\sim 4$ for the echo condition $\tauJM{1}=\tauJM{2}$. This modest enhancement is consistent with decoupling a portion of the noise from the environment.  A model of classical, slowly fluctuating hyperfine field gradients [Fig.~\ref{FigMiniEcho}(d)], yields a value for the standard deviation of $\dBLM$ and $\dBMR$ of 3.4 mT, and indicates these gradients to be the dominant noise source for this pulse sequence.

\vspace{0.15 cm}
\textit{{Tomographic Characterization of the System}}
\vspace{0.15 cm}

Qubit performance is commonly assessed by state and process tomography.  These techniques require a well-characterized set of measurements that give enough information to reconstruct all the matrix elements of the density matrix of the system.  It is important to recognize, however, that measurements themselves are subject to noise, relaxation, and systematic errors, and may not be well described by idealized projective measurements. To accommodate both state preparation and measurement errors we implement a self-content approach that combines measurement data and models of system dynamics within a maximum likelihood estimation routine. Self-consistent tomography along similar lines has been carried out recently for superconducting qubits in Ref.~\cite{2012arXiv1211.0322}.

Consider single-shot measurement in the singlet-triplet basis, for the moment ignoring leakage into state $\Q$.  The singlet measurement fidelity $F_S$ is the probability that the measurement correctly registers the singlet outcome when measuring a system prepared in the singlet state; similarly, $F_T$ is the probability that system prepared in the triplet state yields the triplet outcome for the measurement~\cite{BarthelSingleShot}.  Given that the measurement is defined to be along the $\SL$-$\TL$ or $\SR$-$\TR$ axis of the Bloch sphere, these two numbers completely characterize an imperfect two-outcome measurement on a qubit.  On the other hand, measurements along other directions require rotations which are themselves imperfect as well. This, then, requires a general description of a noisy measurement that includes errors in measurement {\em direction} as well as reduced fidelities.  Such a description is provided by the formalism of Positive Operator-Valued Measure (POVM) elements.  As all our measurements have two outcomes, the POVM describing each measurement basis choice $i$ is given by a single positive Hermitian matrix $\ES{i}$ associated with the ``singlet'' outcome.  (The corresponding ``triplet'' outcome is associated with the matrix $\ET{i}$.)  The eigenvectors of $\ES{i}$ determine the axis of the Bloch sphere along which the measurement is made.  The eigenvalues are bounded between zero and one, with the larger eigenvalue of $\ES{i}$ equal to $F_S$ while the smaller eigenvalue is $1-F_T$.  Using the POVM formalism, the probability that a system described by a density matrix $\rho$ will yield the singlet outcome when the measurement of basis $i$ is performed is
\beq
\label{POVMeqn}
P_i(\rho) = \mathrm{Tr}[\ES{i}\, \rho].
\eeq

The POVM formalism can be applied to the three-state system of qubit states plus leakage state $\Q$, in which case the $\ES{i}$ are $3 \times 3$ Hermitian matrices. As our measurements are insensitive to coherence between the leakage and qubit states, which in any case is expected to be small, we restrict $\ES{i}$ to be incoherent with the $\Q$ space (i.e., each $\ES{i}$ has support on the reduced qubit subspace together with a population in $\Q$).  We note that the $\Q$ state will, with an idealized spin-to-charge measurement, always yield the triplet outcome. As such, the $\Q$-population of $\ES{i}$ quantifies the error-induced probability that a system prepared in the $\Q$ state will instead yield the singlet outcome.

\vspace{0.15 cm}
\textit{Measurement Tomography}
\vspace{0.15 cm}

Measurement of the four matrices, $\ES{i}$, $i$=1--4, is required for state tomography, additionally yielding the population of the leakage state.  Each of the four $\ES{i}$ has five parameters associated with the qubit state and the leakage population, for a total of 20 unknown quantities~\cite{POVMfootnote}.  Measurement tomography therefore requires measurement outcome statistics on five well-characterized input states, $\rho_{j}$, using each of the four generalized measurements to yield 20 independent observed probabilities, $P_{ij}$.  We then solve $P_{ij} = \mathrm{Tr}[\ES{i}\rho_{j}]$ for $\ES{i}$ self-consistently, subject to the constraints on the eigenvalues of $\ES{i}$ to lie between 0 and 1. Details are given in the Methods section and the Supplementary Information.

\begin{figure*}[ht!]
\includegraphics{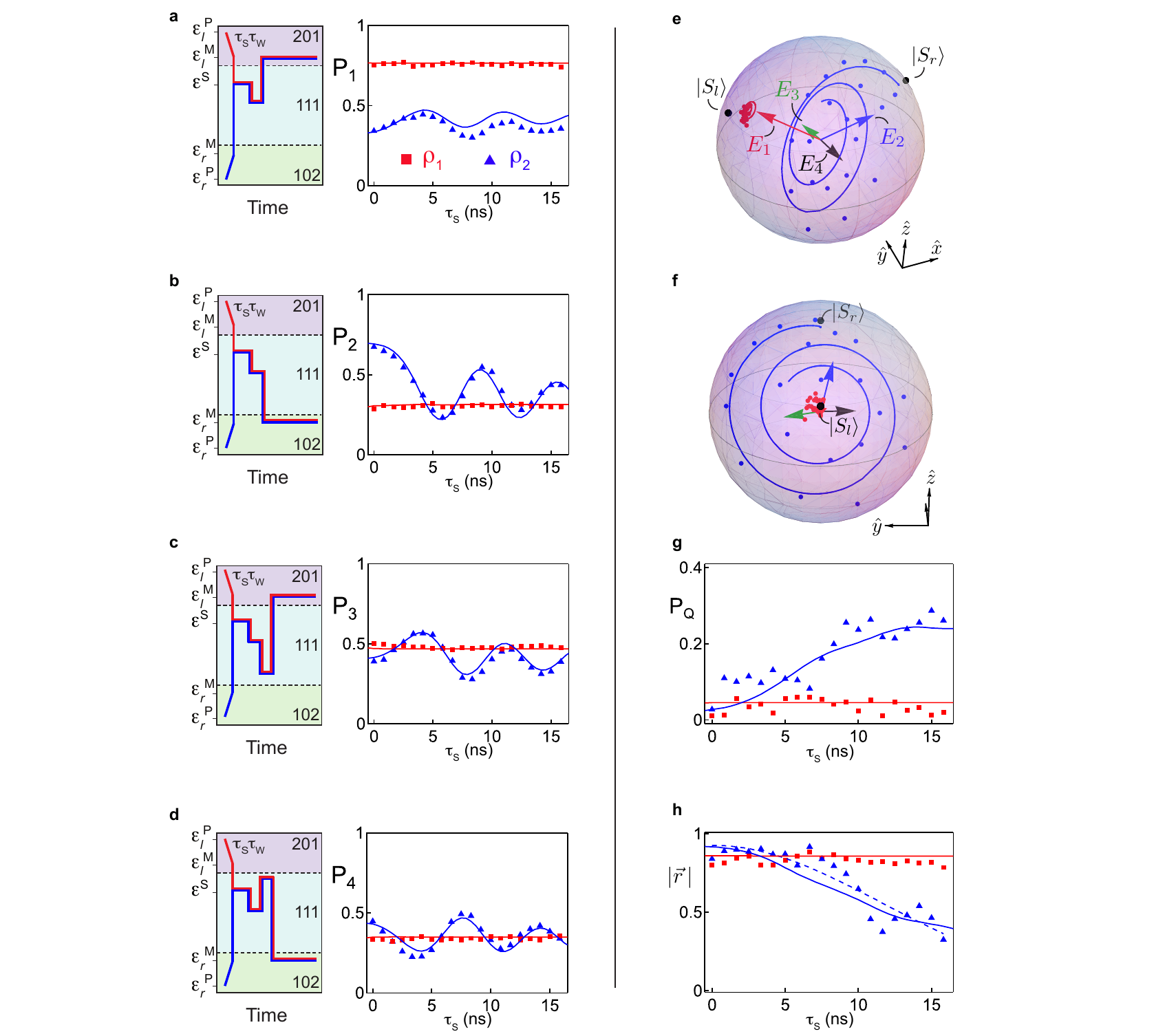}%[width = 2.5 in]Information
\caption[Measurement and State Tomography]{\label{FigBloch}{\bf Measurement and State Tomography.}
~State tomography of the qubit during evolution around $\JL$. Measurement outcome probabilities for four measurement bases for initial states $\rho_1$ (red) and $\rho_2$ (blue), pulsed towards $\eps{S}{}$ near the 201-111 charge transition, producing rotation mostly around $\JL$. The solid curves in (a-d) are 
a fit to the model (see Supplementary Information). Solid curves in (g) and (h) are generated from the model in (a-d). (a) Schematic and measurements of $P_{1}(\rho_1)$ (red) and $P_{1}(\rho_2)$ (blue); (b) $P_{2}(\rho_1)$ (red) and $P_{2}(\rho_2)$ (blue); (c)$P_{3}(\rho_1)$ (red) and $P_{3}(\rho_2)$ (blue); (d) $P_{4}(\rho_1)$ (red) and $P_{4}(\rho_2)$(blue).  (e-f) Views of Bloch sphere with measurement axes. Graphical representation of the qubit portion of $\ES{1}$ (red), $\ES{2}$ (blue), $\ES{3}$ (green) and $\ES{4}$ (black). (g) Population of the leakage state as a function of separation time.  (h) The length of the Bloch vector as a function of separation time data (blue triangles), model (blue curve), and gaussian fit [width $16.4 \pm 0.9$ ns] (dashed blue curve).}
\end{figure*}

Using the reconstructed $\ES{i}$, we extract the measurement bases and fidelities from the eigenvalues and eigenvectors as discussed above. The singlet outcome fidelity is indicated by the length of the $\ES{i}$ arrow in Figs.~\ref{FigBloch}(e,f). The smaller eigenvalue in the qubit space relates to the fidelity of the triplet outcome, $F_{Ti} = 1- \lambda_{i2}$, while the eigenvalue in the leakage subspace relates to the probability that $\Q$ will have a triplet outcome, $F_{Qi} = 1-\lambda_{i3}$. Finally, the measurement visibility in the qubit subspace is found as $V_i = F_{Si} + F_{Ti}-1= \lambda_{i1} - \lambda_{i2}$.  The average singlet fidelity over all four generalized measurements was found to be 69\%, while the average triplet fidelity was 80\%, giving an overall average measurement fidelity in the qubit subspace of 75\% with an average measurement visibility of 49\%. 

\vspace{0.15 cm}
\textit{State Tomography}
\vspace{0.15 cm}

We can now perform state tomography on arbitrary states of our system using our set of tomographically characterized generalized measurements despite the fact that these measurements are inherently noisy. As a demonstration, we generated sets of unknown states by performing a simple rotation around $\sigL$ in the presence of dephasing for multiple fresh input states. In Fig.~\ref{FigBloch}, two separate input states, $\rho_1$ (red) and $\rho_2$ (blue), are prepared and then pulsed to a negative detuning where $\JL\gg\JR$, followed by a generalized measurement (the markers in Fig.~\ref{FigBloch}(a-d)). Using our descriptions of $\ES{i}$, we are able to reconstruct the state at a set of time intervals during the evolution by solving equation \eqref{POVMeqn} again for $\rho$ (the markers in Fig.~\ref{FigBloch}(e,f)). The theory curves overlaid on top of the data in Fig.~\ref{FigBloch}(a-h) are fits to the Liouville-von Neumann equation,  
accounting for the finite bandwidth of the coax and a theoretical model of the exchange profile, in the presence of the nuclear noise determined from the calibration procedure. 

As expected, pulsing towards a negative detuning yields an outcome that depends on the input state. For the states prepared as $\rho_1 = \rhoL$, the red markers and curves in Fig.~\ref{FigBloch}(e,f), sitting at a position of large $\JL$ only imparts a trivial phase. At this detuning, $\SL$ is split off energetically from $\Q$, suppressing leakage out of the qubit space. This is observed on the Bloch sphere as a collection of points near the idealized $\SL$ state.

The state that is prepared as $\rho_2=\rhoR$, the blue markers and curves in Fig.~\ref{FigBloch}(e,f), has a very different response. Since it is an eigenstate of $\JR$, at this detuning it is a superposition of the ground and excited qubit states. As a result, it precesses around the $\sigL$ axis in the presence of dephasing, which causes the state to spiral inwards towards the rotation axis.

At this detuning a fraction of $\rho_2$ is in the excited qubit state, which is energetically close to the $\Q$ state, allowing the Overhauser gradients to rotate that fraction out of the qubit space. This leakage into the $\Q$ state occurs on a 10 ns timescale in both the data and the theory, which increases the decay of the Bloch vector $\vec{r}$ towards the center of the Bloch sphere in Figs~\ref{FigBloch}(e,f,h) as the probability exits the qubit subspace. The leakage is seen clearly in the rise of the $\Q$ population, $P_Q$, in Fig.~\ref{FigBloch}(g). The qubit state vector length in Fig.~\ref{FigBloch}(h) acts as a measure of qubit coherence and population, and it decays with a $T_2^*\sim16$ ns, which is consistent with previous measurements.

In conclusion, we have demonstrated initialization, two-axis electrical control, and self-consistent state reconstruction of an exchange-only spin qubit. The exchange interaction allows extremely fast qubit operation. The method of tomographic calibration we developed can be applied directly to other qubit systems to quantify measurement errors. Future work will include investigating regimes where $\JL$ and $\JR$ are simultaneously much larger than the nuclear gradient Zeeman energy, which would suppress leakage into $\Q$, and structures comprising of six dots that implement a fast two-qubit gate \cite{divincenzo2000universal}.

%\begin{acknowledgments}
\textit{Acknowledgements}.---We gratefully acknowledge support from IARPA through the MQCO program, the Danish National Research Foundation, and the Villum Foundation. S.B.~and A.D.~acknowledge support from the ARC via the Centre of Excellence in
Engineered Quantum Systems (EQuS), project number CE110001013. We thank Oliver Dial, Bert Halperin, Ferdinand Kuemmeth, Thaddeus Ladd, and Amir Yacoby for useful discussions. We thank Brandon Armstrong for technical contributions.
%\end{acknowledgments}

\vspace{0.15cm}
\textit{Methods}
\vspace{0.15cm}

Measurement tomography requires a choice of five input states, $\rho_j$, which span the qubit subspace. The initialization states $\SL$ and $\SR$ provide $\rho_1$ and $\rho_2$ respectively. We create two additional states $\rho_3$ and $\rho_4$ by rotating $\SL$ around $\sigR$ at $\eps{S}{3}$ and $\SR$ around $\sigL$ at $\eps{S}{4}$ respectively. These four input states span the qubit space. The rotated states are subject to rotation errors, dephasing, and leakage, which we need to characterize using a phenomenological model for the dynamics and fit the parameters of this model using experimental data. 

To facilitate this, we use a series of states for $\rho_3$ and $\rho_4$. The noisy evolution of $\rho_3$ and $\rho_4$ is then modeled with a generalization of $\HJ{}$ to the larger manifold of $\TM$, $\SM$, and $\Q$, including the effects of $\dBLM$ and $\dBMR$ (see Supplementary Information for details). During the calibration of the $\ES{i}$'s, we compare the model of this evolution to the series of states $\rho_3(\tauJM{3})$ created by initializing $\SL$ and rotating around $\JR$ at $\eps{S}{3}$ for a set of times $\tauJM{3}$ before measuring. The series of states $\rho_4(\tauJM{4})$ was produced in a similar fashion by preparing $\SR$ and rotating for a set of times $\tauJM{4}$ about $\JL$ at $\eps{S}{4}$. These series of states contain enough information to determine the strength of the nuclear dephasing and the exchange axes at $\eps{S}{3}$ and $\eps{S}{4}$.

The final input state, $\rho_5$, is chosen to be a completely mixed state with no coherences remaining and a significant weight in the leakage state. 
This choice allows for accurate measurements of the $\rhoQQ$ parameter in each $\ES{i}$. The ensemble of $\rho_5$ was prepared by performing repeated pulses to dephase around $\sigL$ and $\sigR$ over a distribution of $\dBLM$ and $\dBMR$. 

With the observed statistics $P_{ij}$ for five known input states $j$ using four measurements $i$, we determine $\ES{i}$ by fitting the calibration probabilities from all of our input states to our model of the noisy evolution to produce a Maximum Likelihood Estimate (MLE) for $\ES{1}$-$\ES{4}$ as well as the nuclear noise and exchange during the calibration~\cite{POVMnoiseFootnote}. For the data in Fig.~\ref{FigBloch}, the standard deviations of $\dBLM$ and $\dBMR$ were $\sim 2.5$ mT, which is consistent with the earlier estimations extracted from the FID and echo data.

\setcounter{figure}{0}
\setcounter{equation}{0}
\renewcommand{\thefigure}{S\arabic{figure}}  
\renewcommand{\theequation}{S\arabic{equation}} 
\renewcommand{\figurename}{Figure} 
\renewcommand{\thefootnote}{S\arabic{reference}} 

\pagebreak
\onecolumngrid
\pagebreak
\section*{Supplementary Information for Self-Consistent Measurement and State Tomography of an Exchange-Only Spin Qubit}

This supporting document describes further details of the fabrication, state readout, noise modeling, and measurement tomography techniques. The measurement tomography section details the pulse sequences and fitting routines used to extract the POVM elements, as well as the effects of finite bandwidth limitations on the state reconstruction.

\subsection{Device}
The three-electron system was confined in a lateral triple quantum dot formed in the two-dimensional electron gas (2DEG) at the GaAs/Al$_{0.3}$Ga$_{0.7}$As interface 110 nm below the surface of the heterostructure.
The GaAs/Al$_{0.3}$Ga$_{0.7}$As heterostructure was grown on a solid-source Varian Gen II molecular beam epitaxy (MBE) system equipped with an arsenic valved cracker source to provided As$^2$ for the growth. The heterostructure was grown on a semi-insulating (100) GaAs substrate with a growth rate of ~1 $\mu$m/hr. The 2DEG is formed by a Si modulation doping ($\delta$-doping) of $\sim$4 $\times$ 10$^{16}$ m$^{-2}$ (40 nm away from the 2DEG interface). Hall effect measurement done at 20K gives a 2DEG density of $\sim$2.6 $\times$ 10$^{15}$ m$^{-2}$ and a mobility of $\sim$43 m$^{2}$/V s.

 High bandwidth coaxial lines were attached to the left, middle, and right plunger gates of the triple quantum dot, and a radio-frequency (rf) reflectometry circuit was connected to a neighboring quantum dot for fast state readout\cite{Reilly_APL07,BarthelSingleShot}. The experiment was performed in a dilution refrigerator equipped with a cryogenic amplifier (noise temperature $T_N \sim 3$ K), with an electron temperature of $\sim$ 120 mK. An in-plane external magnetic field of 300 mT was applied along the dot connection axis [see Fig~1(a) in main text].

\subsection{Measurement and Normalization}\label{norm}
\subsubsection{Normalizations Based on Single-Shot Outcomes}
A uniform normalization procedure was used for all data in Figs.~2-5 to convert the measured reflectometry signals into output probabilities. For a given set of pulse parameters ($\eps{S}{}$, $\tau_{\rm{S}}$, $\tauJM{1}$+$\tauJM{2}$, $\tauJM{1}$-$\tauJM{2}$, etc.),  the qubit was measured using four preparation and measurement routines. In the first two routines, the state $\SL$ was prepared in 201 then measured either in 201 ($\SL$ projection, yielding $P_1$) or in 102 ($\SR$ projection, yielding $P_2$). In the other two routines, the state $\SR$ was prepared in 102 then measured either in 201 ($\SL$ projection) or in 102 ($\SR$ projection). Each measurement consisted of sitting at the measurement point---$\eps{M}{l}$ ($\eps{M}{r}$) for $\SL$ ($\SR$) readout---and integrating the demodulated rf signal reflected from the impedance transforming circuit \cite{Reilly_APL07} attached to the rf-sensor quantum dot for $\tauJM{M}=50~\mu$s to yield $\vrf $.

This process was then repeated while stepping one of the pulse parameters ($\eps{S}{}$, $\tau_{\rm{S}}$, $\tauJM{1}$+$\tauJM{2}$, $\tauJM{1}$-$\tauJM{2}$, etc.). Each sequence was then repeated $2^{13}$ or $2^{14}$ times to obtain measurement statistics. The resulting data was then histogrammed, following the procedure in Ref.~\cite{BarthelSingleShot}, and fit to a function of the form,

\begin{figure}[h]
\includegraphics{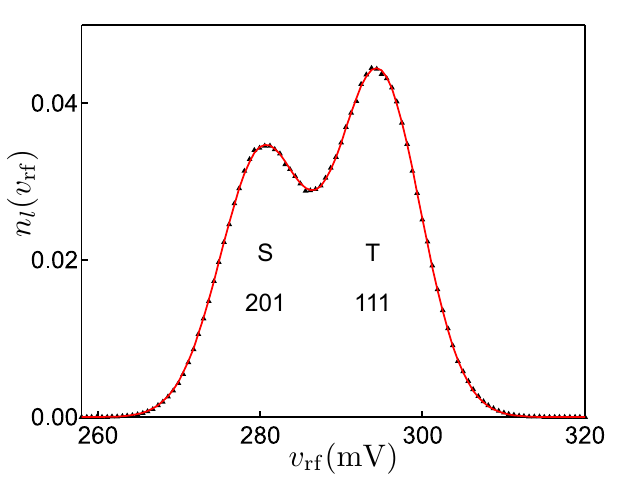} %[width = 3.2 in]
\caption[Single-shot histogram]{
\label{FigHist}{\bf Single-shot histogram.}
~A histogram of outcomes for Fig.~3(a) in the main text. Red solid curve is a fit to equation \eqref{eq:Hist}.}
\end{figure}
\begin{align}\label{eq:Hist}
 	n_{l}(\vrf) &= \frac{P_{1}}{\sqrt{2 \pi \sigma^2}}\,\exp\left[-\frac{(\vrf-\vrf^{201})^2}{2\sigma^2}\right]\nonumber\\
& + e^{-\tauJM{M}/T_1}\frac{(1-P_{1})}{\sqrt{2 \pi \sigma^2}}\,\exp\left[-\frac{(\vrf-\vrf^{111})^2}{2\sigma^2}\right]\nonumber\\
& +\int_{\vrf^{201}}^{\vrf^{111}}\frac{d\rm{V}}{\sqrt{2 \pi \sigma^2}}\frac{\tauJM{M}}{T_1}\frac{(1-P_{1})}{\Delta\vrf}\exp\left[-\frac{\tauJM{M}}{T_1}\frac{\rm{V}-\vrf^{201}}{\Delta\vrf}-\frac{(\vrf-\rm{V})^2}{2\sigma^2}\right],
\end{align}
where $n_{l}(\vrf)$ is the fraction of histogram events with outcomes $\vrf$ for a measurement in 201, $\vrf^{201}$ is the reflected voltage corresponding to double occupancy in the left dot, $\vrf^{111}$ is the reflected voltage corresponding to single charge occupancy in the all three dots,  $\Delta \vrf \equiv \vrf^{111}-\vrf^{201}$, $P_{1}$ is the fraction of 201 outcomes in the data set, $T_1$ is the relaxation time at $\eps{M}{l}$, $\tauJM{M}$ is the measurement time, and $\sigma$ is the standard deviation of the histogram peaks due to noise in the rf equipment and shot noise intrinsic to the rf sensor dot. For measurements in the right dot, $n_{r}$ has an identical form, with all 201 notations replaced with 102 and $P_{1}$ replaced with $P_2$.

The extracted parameters $\vrf^{201}$ and $\vrf^{111}$ are then used to normalize the return probabilities on the left side,
\beq\label{eq:VrftoP}
	P^0_1 (\eps{S}{},\tauJM{S}) = \frac{\langle \vrf (\eps{S}{},\tauJM{S})\rangle - \vrf^{111}}{\vrf^{201}-\vrf^{111}},
\eeq
where $\langle \vrf (\eps{S}{},\tauJM{S})\rangle$ is the average voltage over all repetitions of the measurement sequence for a specific $\eps{S}{}$ and $\tauJM{S}$. $P^0_2$ is normalized similarly, with $\vrf^{201}$ replaced by $\vrf^{102}$.

Equation \eqref{eq:VrftoP} converts $\vrf$ into a probability, but it does not account for relaxation during the measurement time $\tauJM{M}$, where a 111 state relaxes to a 201 state for $\SL$ projections or a 102 state for $\SR$ projections. Relaxation during the measurement was accounted for using a two step process. The histogram shape is only weakly dependent on the precise value of $T_1$, but failing to allow for relaxation of the $111$ charge state would underestimate the separation between histogram peaks. $T_1$ decay was therefore included in equation \eqref{eq:Hist} to determine the peak positions $\vrf^{201}$ and $\vrf^{111}$ accurately, but the $T_1$ fit parameter is not itself an accurate measurement of the relaxation time in the data. In order to more accurately correct for relaxation, we project a state prepared as $\SR$ in 102 against $\SL$ in 201, and record the probability as $P^{cal}_1$, at the beginning of each sequence and compare it with the theoretical value $|\langle S_r | S_l\rangle|^2 = 0.25$; we confirm the theoretical value by measuring relaxation as a function of measurement time $\tauJM{M}$, as described in Sec.~B.2. Traces were then corrected as
\begin{align}%\eta = 1-\frac{P_{1}^0}{1-P_{1}(\rhoR)},
%	1-P_1 &= (1-P^0_1)\frac{1-0.25}{1-P^{cal}_1}\nonumber\\
	P_1 &= 1-(1-P^0_1)\frac{1-0.25}{1-P^{cal}_1}
\end{align}
A measurement of a state prepared as $\SL$ and measured in 102 is similarly used to correct for $T_1$ decay in $P_2$.

\subsubsection{$T_1$ Relaxation During Measurement}
\begin{figure}
\includegraphics{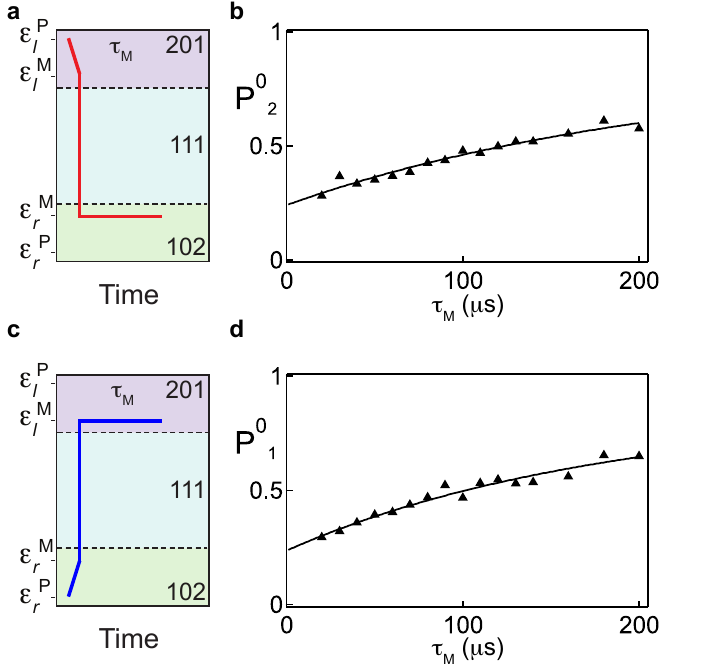} %[width = 3.2 in]
\caption[Measurement relaxation and confirmation of 120$^\circ$ axis separation]{
\label{FigAngleConfirm}{\bf Measurement relaxation and confirmation of 120$^\circ$ axis separation}
~(a) Schematic for preparing $\SL$ and projecting it onto $\SR$. (b) Experiment (triangles) and numerics (solid curve) for the probability of measuring a singlet in 102 if the state was prepared as $\SL$ in 201, as a function of measurement time $\tauJM{M}$. The numerical results are uncorrected for relaxation during the measurement. Numerical result is a fit to the function \eqref{AlexMod} with fit parameters $T_1 = 137\pm9~\mu$s and $|\langle S_r | S_l\rangle|^2 = 0.24 \pm 0.01$ at $\tauJM{M}=0$. (c) Schematic for preparing $\SR$ and projecting it onto $\SL$. Experiment (triangles) and numerics (solid curve) for the probability of measuring a singlet in 201 if the state was prepared as $\SR$ in 102, as a function of measurement time $\tauJM{M}$, uncorrected for relaxation during the measurement. Numerical result is a fit to the function \eqref{AlexMod} with fit parameters $T_1 = 110\pm8~\mu$s and $|\langle S_r | S_l\rangle|^2 = 0.24 \pm 0.02$ at $\tauJM{M}=0$. }
\end{figure}
The relaxation time was extracted from a measurements of the dependence of the uncorrected singlet probabilities $P^0_1$ and $P^0_2$ as functions of the integration time $\tauJM{M}$ for $\SL$ measured in 102 and $\SR$ measured in 201. The observed dependence was well described by exponential relaxation of triplets integrated over the measurement time,
\begin{align}\label{AlexMod}	
	1-P^0_1 &  = \frac{1}{\tauJM{M}}\int_0^{\tauJM{M}}\mathrm{d}t (1-A) e^{-t/T_1}\nonumber\\
		P^0_1 &= 1-\frac{(1-A)T_1}{\tauJM{M}}(1-e^{-\tauJM{M}/T_1}),
\end{align}
where $A$ is the fraction of singlets present at $\tauJM{M}=0$, and corresponds to $|\langle S_r | S_l\rangle|^2$ in the absence of any evolution when pulsing the prepared state to the measurement position. The expression for $P^0_2$ is identical, with $T_1$ referring to the measurement relaxation on the right rather than the left. In the data shown in Fig.~\ref{FigAngleConfirm}, extrapolation to zero measurement time yields $A = 0.24 \pm 0.02$ for both $P^0_1$ and $P^0_2$. This allows us to correct for relaxation by normalizing the data with the theoretical singlet probability.

\subsection{Figure 4b Theory Curves}
\subsubsection{Pure Electrical Dephasing}
We model exchange rotation in the presence of low-frequency detuning noise as:
\beq
	P_1(\tauJM{S}) = \frac{5}{8}\left\{ 1-\cos[\tauJM{S}J(\tauJM{S})] \exp\left[-\left(\frac{\tauJM{S}}{T_2^*}\right)^2\right] \right\},
\eeq
where $T_2^*$ is a characteristic time for dephasing due to electrical noise. Rise-time effects, due for instance to bandwidth limits of the coaxial cable, are modeled as an exponential rise in the exchange,
\beq
	J(t) = J_0\left[1-\exp\left(-\frac{t}{\tau_R}\right)\right].
\eeq
The dashed gray curve in Fig.~4b of the main text used the parameters $J_0 =391$ neV,   $T_2^* = 25$ ns, and $\tau_R = 13$ ns.
\subsubsection{Dephasing due to Nuclei - Ladd Curve}
Following Ref.~\cite{Ladd12}, Eqs.~9, 14-16, we model the effect of nuclear fluctuations on the qubit as
\beq
	P_1(\tauJM{S}) = \frac{1}{2} + \frac{1}{4}\left\{-I_1(\tauJM{S}) + \exp\left[-\frac{3}{2} (\sigma_{\Delta B} \tauJM{S})^2\right]I_2(\tauJM{S})  \right\},
\eeq	
where
\begin{subequations}
\begin{align}
	I_1(t) &= \frac{\sqrt{\pi} J(t)}{4 \sigma_{\Delta B}} \exp\left[\frac{{J(t)}^2}{8 \sigma_{\Delta B}^2}\right]\, \textrm{erfc}\left[\frac{J(t)}{2\sqrt{2}\sigma_{\Delta B}}\right]\times\nonumber\\
	&\left\{ 1-\sqrt{A(t,J,0)}\cos\left[t J(t)+ \frac{1}{2}\cos^{-1}\bigg(A(t,J,0) \bigg)  \right]\right\}\\
	I_2(t) &= \sqrt{A(t,2J,0)}\left\{ \cos[t J(t)]+\cos\left[t J(t)+\frac{1}{2}\cos^{-1}\bigg( A(t,2J,0) \bigg) \right]\right\}\\
	A(t,\xi,\omega) &=\frac{1}{\sqrt{1+\left[\frac{4\sigma_{\Delta B}^2 t}{\xi}(1+2 \omega)\right]^2}}\\
	J(t) &= J_0\left[1-\exp\left(-\frac{t}{\tau_R}\right)\right]
\end{align}
Here, we take the standard deviation of the nuclear fluctuations to be the same in each dot, which simplifies the expressions in Ref.~\cite{Ladd12}. The solid gray curve in Fig.~4b of the main text uses fit parameters $J_0 =386\pm 2$ neV,   $\sigma_{\Delta B} = 1.9\pm 0.2$ mT, and $\tau_R = 13$ ns.
\end{subequations}

\subsubsection{Numerical Model: Electrical and Nuclear dephasing}
In order to incorporate both electrical and magnetic sources of noise, we used a numerical model for the time evolution of an initial $\SL$ in the presence of both exchange interactions $\JL$ and $\JR$, as well as longitudinal field gradients $\dBLM$ and $\dBMR$. The finite bandwidth of the coax and function generator are accounted for with an exponential turn-on as described in Sec.~C.1. We take $\JR$ to be Gaussian distributed, appropriate for small amplitude fluctuations in $\eps{S}{}$ over a range where $\JR$ varies approximately linearly with $\eps{S}{}$, that is, $\delta J \approx (\Delta\JR/\Delta\eps{S}{}) \delta \eps{S}{}$. In the region of detuning where $\textrm{d}\JR/\textrm{d}\eps{S}{} \gg \textrm{d}\JL/\textrm{d}\eps{S}{}$, only fluctuations in $\JR$ were taken into account.  Since the left exchange was decreasing while the right exchange was increasing, they were approximated as:
\begin{align}
	\JL &= \JL^0 \,e^{-t/\tauJM{R}}\\
	\JR &= \JR^0(1- \,e^{-t/\tauJM{R}})+\delta J
\end{align}

The effects of the slowly fluctuating nuclear bath were incorporated by taking an ensemble average over Gaussian distributions of nuclear gradients between the left and middle ($\dBLM$) and middle and right ($\dBMR$) dots. Limiting this model to detuning regions away from the $\Qp$-$\SL$ and $\Qp$-$\SR$ anticrossings, transverse components of the hyperfine field can be safely neglected, leaving only gradients between longitudinal components. This model assumes a  Gaussian distribution of classical nuclear gradients, with no back-action on the nuclei from the qubit. 

Explicitly, $P_{1}$ was evaluated numerically using a uniform step size,
\begin{align}
	P_{1}(\tauJM{S}) &= \int \frac{\textrm{d}\dBLM\, \textrm{d}\dBMR \textrm{d}\delta J}{(2\pi)^{3/2}\sigma_B^2 \sigma_J} \left |\langle S_l |e^{-i\h{H}_{1}\tauJM{S}/\hbar}\SL \right|^2 e^{-(\dBLM^2+\dBMR^2)/(2\sigma_B^2)-(\delta J)^2/(2\sigma_J^2)}\\
	 &\approx \sum \frac{(\Delta B_{max} - \Delta B_{min} )^2(J_{max} - J_{min} )}{(2\pi)^{3/2}\sigma_B^2\sigma_J N_{step}^3} \left |\langle S_l |e^{-i\h{H}_{1}\tauJM{S}/\hbar}\SL \right|^2 e^{-(\dBLM^2+\dBMR^2)/(2\sigma_B^2)-(\JR-\JR^0)^2/(2\sigma_J^2)},
\end{align}
where $\tauJM{S}$ is the time spent during the rotation, $\sigma_B$ is the standard deviation of the Gaussian distribution of classical values that each nuclear gradient could achieve, $\sigma_J$ is the standard deviation of the Gaussian distribution of $\JR$ values, $N_{step}$ is the number of discrete values sampled for each Gaussian, $\Delta B_{max}$ and $\Delta B_{min}$ are the limits of $\dBLM$ and $\dBMR$ values sampled, $J_{max}$ and $J_{min}$ are the limits of $\delta J$ values sampled, $\tauJM{R}$ is the turn-on time for the exchange, $\h{H}_1$ is the Hamiltonian at the dephasing position. The Hamiltonian consisted of two parts, the model laid out in Ref~\cite{LairdTdot}, and a nuclear Hamiltonian; $\Heff{1}{} = \h{H}_J(\JL,\JR)+\gamma\h{H}_B(\dBLM,\dBMR)$, where $\gamma= g\mu_B = -25.4 \,\rm{neV}/\rm{mT}$.

We can write the exchange Hamiltonian in the basis of ($\TM$-$\SM$-$\Q$) as:
\begin{align}\label{ExchangeHamiltonian}
	\h{H}_{J} &=
\begin{pmatrix}	
	-\frac{3}{4}(\JL+\JR) & \frac{\sqrt{3}}{4}(\JL-\JR)&0\\
	\frac{\sqrt{3}}{4}(\JL-\JR) & -\frac{1}{4}(\JL+\JR)&0\\
	0&0&0\\ 
\end{pmatrix}	
\end{align}
Here, the zero energy state has been shifted relative to $\HJ{}$ in the main text to make the energy of the $\Q$ state zero at zero detuning. This brings our notation into agreement with Ref.~\cite{LairdTdot}.

The longitudinal nuclear terms in this basis are:
\begin{align}\label{NuclearHamiltonian}
	\h{H}_{B} &= 
\begin{pmatrix}	
	\frac{1}{6}(\dBLM-\dBMR) & \frac{1}{2\sqrt{3}}(\dBLM+\dBMR)&-\frac{1}{3\sqrt{2}}(\dBLM-\dBMR) \\
	 \frac{1}{2\sqrt{3}}(\dBLM+\dBMR)&-\frac{1}{6}(\dBLM-\dBMR)&\sqrt{\frac{1}{6}}(\dBLM+\dBMR)\\
	-\frac{1}{3\sqrt{2}}(\dBLM-\dBMR)&\sqrt{\frac{1}{6}}(\dBLM+\dBMR)&0\\ 
\end{pmatrix}	
\end{align}
where $\dBLM = (B^z_{1}-B^z_{2})$ and  $\dBMR = (B^z_{2}-B^z_{3})$ are the differences in local magnetic field along the $\hat{z}$ direction. Terms that only contribute a global phase in this basis have been dropped for clarity.

A fit to this model yields: $\JL^0 = 21\pm 8$ neV, $\JR^0 = 388 \pm 2$ neV, $\sigma_B = 2.0\pm0.1$ mT, $\sigma_J = 19 \pm 2$ neV, $\tauJM{R}= 12.6 \pm0.2$ ns. The distributions were each sampled evenly 12 times ($N_{step}$) each for a total of $12^3 = 1728$ samples between $3\sigma_B$ and $-3\sigma_B$ for the nuclei and between $3\sigma_J$ and $-3\sigma_J$ for $\delta J$. This gives the solid black curve in Fig.~4b.

\subsection{Figure 5d Echo with Hyperfine Dephasing and Leakage, without Electrical Noise}

The partial echo (Fig.~5 in the main text) was analyzed using a model similar to the one used in Figure 4c. The timescales involved in the echo are much longer than the rise time $\tauJM{R}$, so the phenomenological exponential turn on  of the exchange is removed for simplicity. In addition, since the dephasing was dominated by nuclei in Figure 4 the noise on $\JR$ is omitted.

The model includes the evolution of an initial $\SL$ under the action of the three exchange pulse sequence in the limit of instantaneous rise times in the qubit environment of 111. The pulses are evaluated in a piecewise-static manner, ignoring the weak adiabatic effects associated with pulsing from one detuning position to the other. 
Explicitly, $P_{1}$ was evaluated numerically using a uniform step size as:
\begin{align}
	P_{1}(\tau_1,\tau_2) &= \int \frac{\rm{d}\dBLM\, \rm{d}\dBMR}{2\pi\sigma_B^2}\, \left|\langle S_l |e^{-i\Heff{1}{}\tauJM{2}/\hbar}\,e^{-i\Heff{2}{}\tauJM{\pi}/\hbar}\,e^{-i\Heff{1}{}\tauJM{1}/\hbar}\SL\right|^2\, e^{-(\dBLM^2+\dBMR^2)/(2\sigma_B^2)}\\
	 &\approx\sum \frac{(\Delta B_{max} - \Delta B_{min} )^2}{2\pi\sigma_B^2N_{step}^2} \, \left|\langle S_l |e^{-i\Heff{1}{}\tauJM{2}/\hbar}\,e^{-i\Heff{2}{}\tauJM{\pi}/\hbar}\,e^{-i\Heff{1}{}\tauJM{1}/\hbar}\SL\right|^2\, e^{-(\dBLM^2+\dBMR^2)/(2\sigma_B^2)},
\end{align}
where $\tauJM{1(2)}$ is the time before (after) the $\pi$ pulse, $\sigma_B$ is the standard deviation of the Gaussian distribution of nuclear gradients, $N_{step}$ is the number of discrete values sampled for each gradient, $\Delta B_{min}$ and $\Delta B_{max}$ are the limits of $\dBLM$ and $\dBMR$ values sampled, $\Heff{1}{}$ is the Hamiltonian at the dephasing position, including nuclei, and $\Heff{2}{}$ is the Hamiltonian for the $\pi$ pulse. The Hamiltonians were of the same form as equations \eqref{ExchangeHamiltonian} and \eqref{NuclearHamiltonian}, with $\Heff{1}{} = \h{H}_J(0,\JR)+\gamma\h{H}_B(\dBLM,\dBMR)$, $\Heff{2}{} = \h{H}_J(\JL,0)+\gamma\h{H}_B(\dBLM,\dBMR)$.
The values used in the model were $\gamma= g\mu_B = -25.4 \,\rm{neV}/\rm{mT}$, $\sigma_B = 1.7\,\rm{mT}$, $\JR = 276 \,\rm{neV}$, $\JL = 824 \,\rm{neV}$. The magnetic field gradients were each sampled uniformly between $3\sigma_B$ and $-3\sigma_B$, with $N_{step} = 40$.

\subsection{Measurement Tomography for the Exchange Only Qubit}
\subsubsection{Measurements and Measurement Operators}
In order to determine the populations ($\rhoZZ$, $\rhoOO$) and coherences ($\reRhoZO$, $\imRhoZO$) in the qubit subspace (four unknowns), as well the population of the leakage state ($\rhoQQ \equiv 1 - \rhoZZ-\rhoOO$), we need to perform four measurements. The measurement probabilities can be expressed in the following fashion:
 \begin{subequations}\label{EqnTomography}
 \begin{alignat}{3}
	P_{1}(\rho)&= \mathrm{Tr} [ \ES{1} \rho ]\\
	P_{2}(\rho)&=\mathrm{Tr} [ \ES{2} \rho ]\\
	P_{3}(\rho)&= \mathrm{Tr} [ \ES{3} \rho ]\\
	P_{4}(\rho)&=\mathrm{Tr} [ \ES{4} \rho ]
\end{alignat}
\end{subequations}
where $\rho$ denotes an unknown input state and $\ES{i}$ is a measurement operator that describes the fidelity of a singlet outcome for a measurement in the $i^{\mathrm{th}}$ basis
If we have a set of five (or more) known input states, $\rho_j$, one can use those states to measure the $\ES{1}$, $\ES{2}$, $\ES{3}$, and $\ES{4}$ by solving the set of equations~\eqref{EqnTomography}. Once the $\ES{i}$ are determined, we can reconstruct any unknown state $\rho$ from the four probabilities, $P_1$, $P_2$, $P_3$, and  $P_4$.

 Our measurement tomography approach uses known input states to characterize the measurement operators $\ES{i}$. Of the five required input states, three are relatively easy to prepare. These are the two initialization states, $\SL$ and $\SR$, and the completely mixed state. Their density matrices in the $\TM$-$\SM$-$\Q$ basis can be written as:
\begin{align}
	\rho_1 &= \rhoL =
\begin{pmatrix}
	\frac{3}{4} & -\frac{\sqrt{3}}{4} & 0\\
	-\frac{\sqrt{3}}{4} &\frac{1}{4} & 0\\ 
	0&0&0\\
\end{pmatrix}\\
	\rho_2 &= \rhoR =
\begin{pmatrix}
	\frac{3}{4} & \frac{\sqrt{3}}{4} & 0\\
	\frac{\sqrt{3}}{4} &\frac{1}{4} & 0\\ 
	0&0&0\\
\end{pmatrix}\\
	\rho_5 &= \ket{mixed}\!\!\bra{mixed} =
\bep
	\frac{1}{3} &0 &0\\
	0& \frac{1}{3}&0\\
	0&0&\frac{1}{3}\\
\eep	
\end{align}

We create $\rho_5$ by pulsing to regions of large $\JL$ and $\JR$ repeatedly, allowing the state to dephase around both rotation axes as well as the nuclear gradients. The preparation of the dephased state is confirmed by comparing measurements of the dephased state that was initially prepared as $\SL$ with the state initially prepared as $\SR$. The only way that these two outcomes will be identical is if they are both completely dephased in the qubit space. The length of the sequence is many times larger than $T^{*}_{2,nuc}$, which when combined with the pulses yields a completely mixed state with $\ket{Q}$ as well\cite{SUPPmixedfootnote}. This state allows for the characterization of the $\Q$ fidelity in each measurements, which is not necessarily identical to the qubit triplet-like ($\TL$ and $\TR$) fidelity. 
 
Those three states are entirely real by construction. It is more challenging to prepare high fidelity states with $\imRhoZO\neq 0$, which are needed to fully characterize the system. These superposition states allow us to characterize the complex quanties of our measurement operators, and in turn allow us to measure any superposition of states in the qubit subspace.
Since we do not have access to initialization states with $\imRhoZO\neq0$, we need to create these states through evolutions under control pulses. These pulses themselves contain noise. The extent to which we correctly account for the dephasing and leakage that occurs during the preparation of these states determines our ability characterize the two rotated input states, $\rho_3 = \rhoThree$ and $\rho_4 = \rhoFour$, and therefore $\ES{3}$ and $\ES{4}$. We incorporate the details of the evolution, including noise, into the estimation of our input states using MLE techniques, which results in a more accurate state reconstruction.

The final algorithm we use simultaneously estimates the evolution of $\rho_3$ and $\rho_4$ as well as the POVM elements $\ES{1}-\ES{4}$ by using a Levenberg-Marquardt least-squares method to minimize the difference between the output of the model and the measured probabilities associated with $\rho_{1}-\rho_{5}$. This gives us our Maximum Likelihood Estimate for $\ES{1}-\ES{4}$.

\subsubsection{Creation and Determination of the Known Input States}

\begin{figure}[h]
\includegraphics{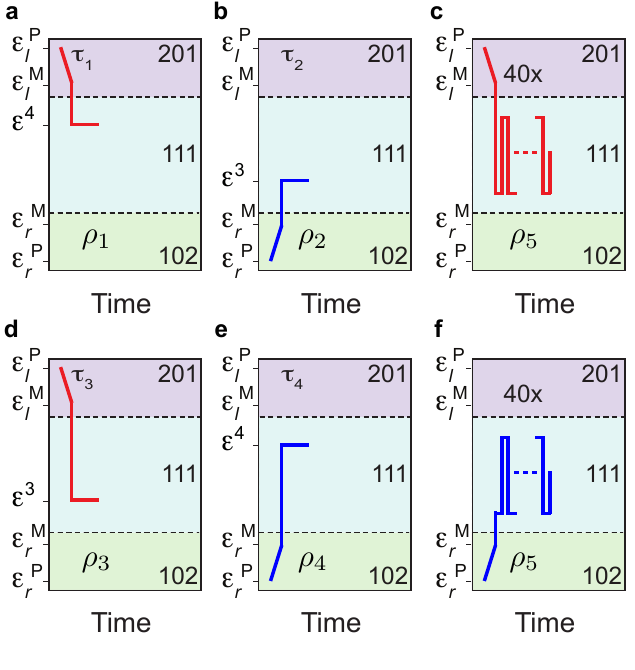}
\caption[Measurement tomography pulse sequences]{
\label{Figrhos}{\bf Measurement tomography pulse sequences.}
~Schematics for pulse sequences that create the five input states. (a) Input state $\rho_1$. (b) Input state $\rho_2$. (c) Input state $\rho_5$ prepared from an initial $\SL$. (d) Input state $\rho_3$. (e) Input state $\rho_4$. (f) Input state $\rho_5$ prepared from an initial $\SR$.}
\end{figure}

One way to minimize the error in estimating $\rho_3$ and $\rho_4$ is to spread $\rho_3$ and $\rho_4$ out into a set of input states that evolved under a common axis of rotation for a range of rotation times $\tauJM{3}$ and $\tauJM{4}$. In other words, $\rho_3$ becomes a set of states $\rho_3(\tauJM{3})$ evolving at $\eps{3}{}$ and $\rho_4$ becomes a set of states $\rho_4(\tauJM{4})$ evolving at $\eps{4}{}$. With these sets of rotation times, we can accurately estimate axis of the rotations and the frequency of the rotation at the fixed detunings $\eps{3}{}$ and $\eps{4}{}$, and therefore each state in the series $\rho_3(\tauJM{3})$ and $\rho_4(\tauJM{4})$. In this manner we can prepare a set of reasonably high fidelity input states that are superpositions of our measurement basis states with complex coefficients.

 Explicitly, we take the effective Hamiltonian for this system, equations \eqref{ExchangeHamiltonian} and \eqref{NuclearHamiltonian}, and evolve the initial density matrix in time. We then average that evolution over a distribution of $\Delta B$'s, calculate the probabilities using equations \eqref{EqnTomography} and compare it to the data. The fitting procedure, if started from good initial guesses (which we generate iteratively through incremental fits, and incremental relaxation of constraints), can generate the MLE for the Hamiltonian parameters and every element of  $\ES{1}$, $\ES{2}$, $\ES{3}$, and $\ES{4}$.

The fit is performed by calculating the theoretical Hamiltonian of the system, using the same techniques that were employed for the FID and the partial echo. Assuming time independent Hamiltonians, $\rho_3(\tauJM{3})$ and $\rho_4(\tauJM{4})$ can be written as
\begin{align}	
	\rho_3(\tauJM{3}) &= \int \frac{\mathrm{d}\dBLM\, \mathrm{d}\dBMR}{2\pi\sigma_B^2}  e^{- i \Heff{3}{} \tauJM{3}} \rho_1 e^{i \Heff{3}{} \tauJM{3}} e^{-(\dBLM^2+\dBMR^2)/(2\sigma_B^2)}\\
	\label{eq:rho4_of_tau}
	\rho_4(\tauJM{4}) &= \int \frac{\mathrm{d}\dBLM\, \mathrm{d}\dBMR}{2\pi\sigma_B^2}  e^{- i \Heff{4}{} \tauJM{4}} \rho_2 e^{i \Heff{4}{} \tauJM{4}} e^{-(\dBLM^2+\dBMR^2)/(2\sigma_B^2)},
\end{align}
where $\Heff{3}{} = \h{H}_J(\JL(\eps{3}{}),\JR(\eps{3}{})) + \gamma\h{H}_B(\dBLM,\dBMR)$, $\Heff{4}{} = \h{H}_J(\JL(\eps{4}{}),\JR(\eps{4}{})) + \gamma\h{H}_B(\dBLM,\dBMR)$, and $\sigma_B$ is the standard deviation of the Gaussian distribution of nuclear gradients. We then generate the probabilities for each measurement using Eqs.~\eqref{EqnTomography}. The Levenberg-Marquardt least-squares algorithm subsequently optimizes the parameters of $\Heff{3}{}$, $\Heff{4}{}$, and $\sigma_B$  to minimize the difference between measured and calculated probabilities. The estimates that the fitting function produce come with error bars, which may be useful indications of the reliability of the MLE output.

There is a further improvement that we can make to this scheme. $\rho_3(\tau_{\mathrm{S}})$ and $\rho_4(\tau_\mathrm{S})$ are formed by pulsing to a region of high exchange, and then pulsing to the settle point, while $\rho_1$ and $\rho_2$ are formed by pulsing from 201 and 102 respectively. This can increase the lowpass effects that the settle point is trying to mitigate. To improve the situation, and to have $\rho_1$, $\rho_2$, and $\rho_5$ have equal weight with $\rho_3(\tau_{\mathrm{S}})$ and $\rho_4(\tau_{\mathrm{S}})$, we can evolve $\rho_1$ and $\rho_2$ under the same Hamiltonians as $\rho_4(\tau_{\mathrm{S}})$ and $\rho_3(\tau_{\mathrm{S}})$ respectively, where $\rho_1$ is approximately an eigenstate of the Hamiltonian that evolved $\rho_4(\tau_{\mathrm{S}})$, and $\rho_2$ is approximately an eigenstate of the Hamiltonian that evolved $\rho_3(\tau_{\mathrm{S}})$.

A few further improvements were made to the calibration routine to mitigate the low-pass effects of the coaxial lines:
\begin{itemize}
\item A voltage overshoot was added to the first 833 ps of the pulse to $\eps{S}{}$ for $\rho_3(\tauJM{3})$ and $\rho_4(\tauJM{4})$. This makes the pulse shapes closer to the ideal square pulse, which is easier to evaluate in our fitting routine.
\item The calibration routine records the evolution of $\rho_3(\tauJM{3})$ and $\rho_4(\tauJM{4})$ for $\tauJM{3},\tauJM{4}>4$ ns, which further reduces the effects of transients at the beginning of the pulses.
\item The settle point employed directly before measurement is a tradeoff between decoupling the rotation pulses from the measurement pulses, which improves with settle time $\tauJM{W}$, and the dephasing brought on by nuclei, which gets worse with increasing $\tauJM{W}$. The settle time needed increases with the amplitude of the pulse directly before the settle point. In order to keep $\tauJM{W}$ to a minimum, $\rho_1$ is placed at $\eps{4}{}$ for $\tauJM{1} = 5$ ns, the evolution point for $\rho_4(\tauJM{4})$, where it is approximately an eigenstate. This prevents the need to pulse all the way from 201 to the settle point, which is a significantly larger amplitude pulse. $\rho_2$ is similarly placed at $\eps{3}{}$ for $\tauJM{2} = 5$ ns to reduce the amplitude necessary to bring it to the settle point as well. The time spent at $\eps{3}{}$ and $\eps{4}{}$ is then incorporated into the fitting routine as well, making $\rho_1(\tauJM{1} = 5\, \textrm{ns})$ and $\rho_2(\tauJM{2} = 5\, \textrm{ns})$.
\item $\rho_1(\tauJM{1} = 5\, \textrm{ns})$, $\rho_2(\tauJM{2} = 5\, \textrm{ns})$, $\rho_5$ are then repeated to give them the same weight in the fitting routine as $\rho_3(\tauJM{3})$ and $\rho_4(\tauJM{4})$.
\end{itemize}

As the last refinement, we allow for small preparation infidelities, which are included as incoherent mixtures of $\TL$+$\Q$ and $\TR$+$\Q$ into $\rho_1(\tauJM{1} = 5\, \textrm{ns})$ and $\rho_2(\tauJM{2} = 5\, \textrm{ns})$ respectively. This adds two more free parameters.

There are 27 unknowns in this problem: five unique quantities for each of the four measurement operators, two unknown exchange terms at $\eps{3}{}$ and two at $\eps{4}{}$, one unknown for the standard deviation of the nuclear gradients, and two unknowns for the preparation infidelities. These 27 parameters were optimized over the entire data set, resulting in the best estimate of the POVM elements and the Hamiltonians at the calibration positions.

Once the system is fully characterized, we invert equation \eqref{EqnTomography}, this time using known $\ES{i}$'s, to solve for the unknown $\rho$. This solution is then used to create the MLE output using the techniques described in Ref.~\cite{PhysRevLett.108.070502}.

\subsubsection{Normalization of Single-Shot Data in Figure 6}
Unlike the charge sensor normalization procedure used for the data in Figs.~2-5, which was covered in Sec.~\ref{norm}, the data in Figure 6 was not normalized by extracting $\vrf^{201}$ and $\vrf^{111}$ and normalizing the average voltage. Instead, we used a single-shot threshold voltage, $\vrf^{T}$, which was chosen to separate 201 and 111 outcomes, as was done in Ref.~\cite{BarthelSingleShot}. This allowed us to make a clean comparison to previous single-shot measurements, including the definitions of measurement fidelity.  The downside of this method is that it reduces the visibility of the oscillations, which is part of the reason for the reduced amplitude in figure 6. $P_i$ is then defined as the fraction of outcomes whose $\vrf$ is on the 201 (for $P_1$ and $P_3$) or 102 (for $P_2$ and $P_4$) side of $\vrf^{T}$.

\begin{figure}[h]
\includegraphics{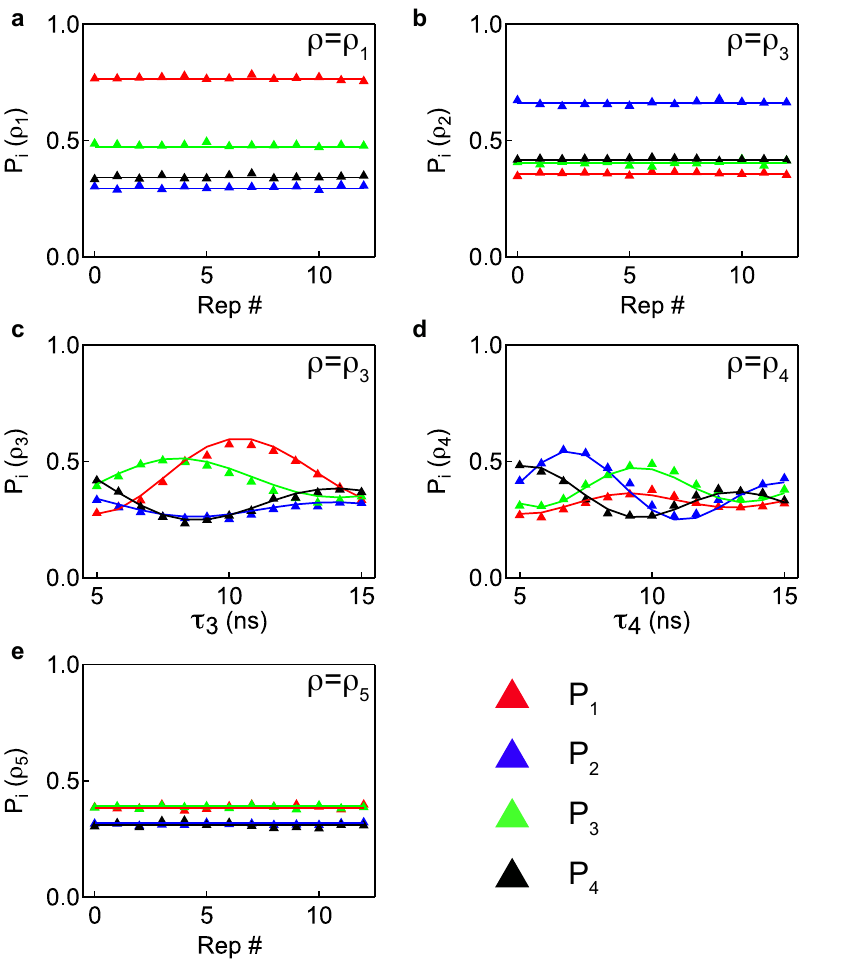} %[width = 3.2 in]
\caption[Measurements of $\rho_j$]{
\label{FigCal}{\bf Measurements of $\rho_j$.}
~The calibration curves and their fits. (a) $\rho_1$ (b) $\rho_2$ (c) $\rho_3$ (d) $\rho_4$ (e) $\rho_5$}
\end{figure}

 Figure \ref{FigCal} shows the example data for the calibration routines which determine $\ES{1}$-$\ES{4}$. The markers are the measurements of the known input states $\rho_{1}$-$\rho{5}$, while the solid curves are the outputs of the MLE routine that extracts the Hamiltonian parameters and the POVM elements. The calibration for $\rho_{1}$, $\rho_{2}$, and $\rho_{5}$ are repeated to provide an equal weighting with $\rho_{3}$ and $\rho_{4}$ in the MLE routine. The Hamiltonian parameters are: for $\rho_3(\tauJM{3})$: $\JL(\eps{3}{}) = 33\pm 8 $ neV, $\JR(\eps{3}{})  = 380\pm 4 $ neV, for $\rho_4(\tauJM{4})$:  $\JL(\eps{4}{}) = 530\pm 8 $ neV, $\JR(\eps{4}{})  = 4\pm 10 $ neV, $\sigma_B = 2.8\pm 0.1 $ mT. 
These parameters define four measurement axes, which when inverted yield the data in Figure 6.

\subsection{Theory Curves in All Panels of Figure 6}
\begin{figure}[!t]
\includegraphics{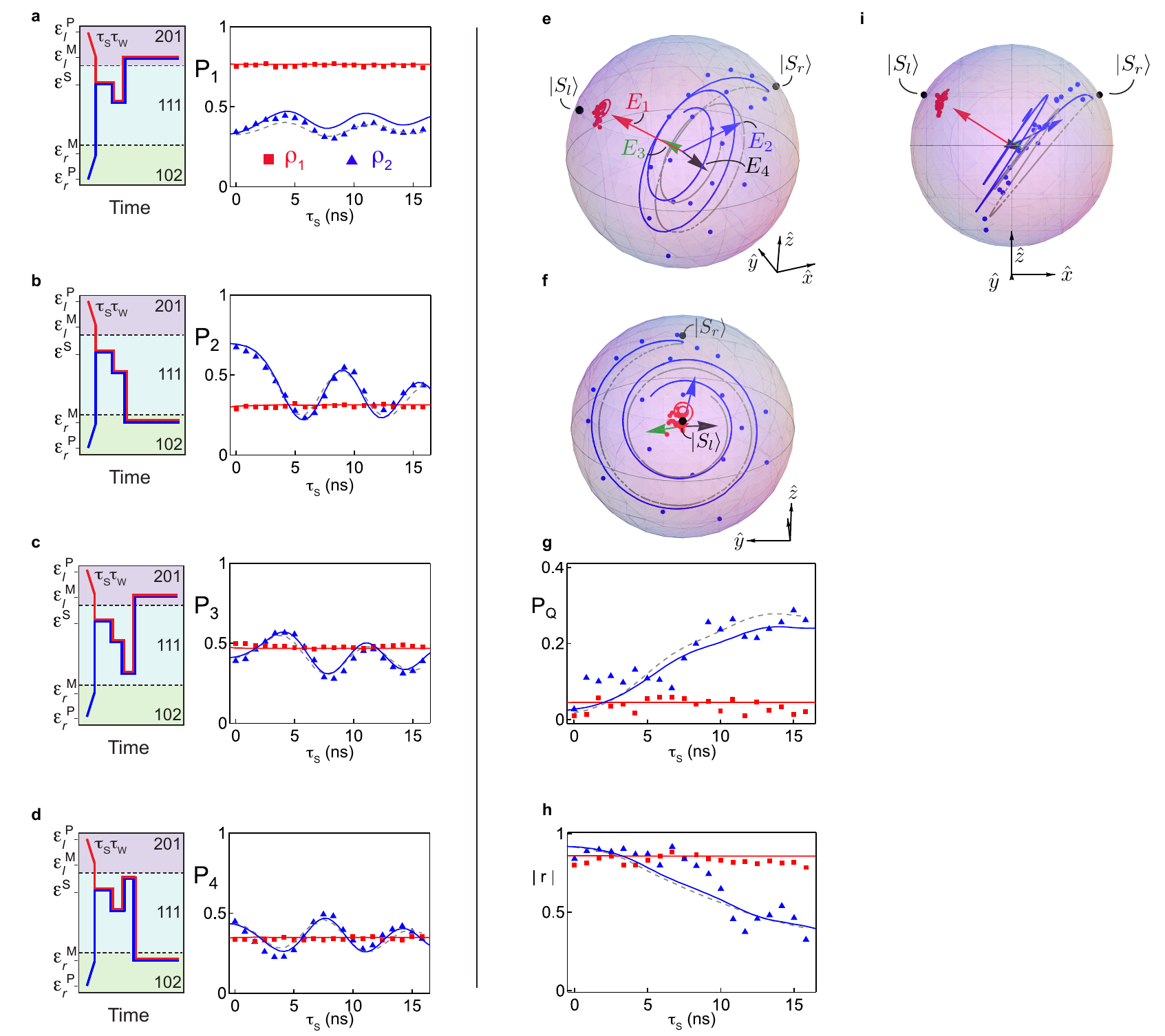} %[width = 3.2 in]
\caption[Tomography with and without bandwidth effects]{
\label{FigHardPulses}{\bf Tomography with and without bandwidth effects.}
~Same data as Fig.~6 of the main paper, here showing a comparison of models using instantaneous pulses (dashed gray), and bandwidth limited pulses (blue).}
\end{figure}
In the presence of sharp (instantaneous) pulses, the theory curve for the $\SR$ initial state stays close to the plane that cuts though $\SR$ and the origin, perpendicular to $\sigL$. This is because the pulses move rapidly through the center of 111, where $\JL$ and $\JR$ are both active. If that region is traversed slowly, the net rotation pulls the state towards $\SL$, with the end result being that subsequent rotations around $\sigL$ trace out a larger circle on the Bloch sphere as it approaches a great circle. To correct for this increased circumference, the full bandwidth limited pulse shape is needed.

The theory curves in Fig.~6 of the main paper were generated by solving the Liouville-von Neumann equation, $i \hbar \dot{\rho} = [\Heff{}{},\rho]$, where $\Heff{}{}$ is a function of the time-dependent pulse detuning, as well as the nuclear gradients, $\dBLM$ and $\dBMR$. $\JL$ and $\JR$ are modeled following Ref.~\cite{LairdTdot}, with the modification that the tunnel couplings are themselves a Gaussian function of the detuning. This modification was necessary to explain the data in Figs.~3 and 6.
\begin{align}
	\JL(t) &= \frac{\alpha}{2} [-\eps{C}{}-\eps{}{}(t)] + \sqrt{\left\{t \exp\left[{-\left(\frac{\eps{C}{}+\eps{}{}(t)}{W_t}\right)^2}\right]\right\}^2+\frac{\alpha^2}{4}[\eps{C}{}+\eps{}{}(t)]^2}\\
	\JR(t) &= \frac{\alpha}{2} [-\eps{C}{}+\eps{}{}(t)] + \sqrt{\left\{t \exp\left[{-\left(\frac{\eps{C}{}-\eps{}{}(t)}{W_t}\right)^2}\right]\right\}^2+\frac{\alpha^2}{4}[\eps{C}{}-\eps{}{}(t)]^2}
\end{align}
Here, the tunnel coupling width $W_t = \eps{}{0}/0.55$ mV, the charge transition detuning $\eps{C}{} = 18$ mV, the lever arm $\alpha = 40\, \mu$eV/mV, and the tunnel coupling $t = 22\, \mu$eV. 

To model the system accurately, the ordinary differential equation (ODE) was solved up to the end of the rotation at $\eps{S}{}$ for each value of $\tauJM{S}$. Those solutions were then used as the initial values in a second solution, where the system was pulsed to the settle point for a time $\tauJM{W}$ before pulsing the the measurement point. The pulses were modeled as the low passed output of the intended piecewise functions,
\beq	
	\eps{}{A}(t) =  
	\begin{cases}
  		 \eps{0}{}+ (\eps{1}{}-\eps{0}{})[1-\exp(-\frac{t-t_0}{\tau_R})] & \text{if } t_0 \leq t <  t_1 \\
   		\left(\eps{0}{} + (\eps{1}{}-\eps{0}{})[1-\exp({-\frac{t_1-t_0}{\tau_R}})]\right)\exp({-\frac{t-t_1}{\tau_R}})  \\ \,\,\,\,\,\,+ \eps{2}{} [1-\exp({-\frac{t-t_1}{\tau_R}})]   & \text{if } t_1 \leq t < \tauJM{S},
 	 \end{cases}	
\eeq
where $\eps{0}{}$ is the measurement point, $\eps{1}{}$ is a resting point at large exchange prior to the rotation, and $\eps{2}{}$ is the rotation point $\eps{S}{}$. The time spent at the resting point starts at $t_0 = -2.5$ ns, while the time at the rotation point starts at $t_1 = 0$ ns. The set of pulses used in the second ODE were
\beq	
	\eps{}{B}(t) =  
	\begin{cases}
     		\eps{}{A}(\tauJM{S})\exp({-\frac{t-\tauJM{S}}{\tau_R}})  & \text{if } \tauJM{S} \leq t < \tauJM{S}+ \tauJM{W},
 	 \end{cases}	
\eeq
which accounted for the approach to the settle point. After this, the POVM elements take effect, describing the rest of the measurement. 

The time evolution is calculated for each configuration of $\dBLM$ and $\dBMR$, and averaged over a Gaussian weighting for each gradient, the standard deviation of which was extracted in the POVM characterization routines described above. No exchange noise was included in this model.
The measurement probabilities associated with the averaged solution of the Liouville-von Neumann equation were then extracted using equations~\eqref{EqnTomography}. The population of the $\Q$ state was extracted as $P_Q = \rhoQQ$, The Bloch sphere vector length is extracted from the qubit subspace of the density matrix as $|\vec{r}| = \sqrt{(\rho^*_x)^2+(\rho^*_y)^2+(\rho^*_z)^2}$, where 
\beq
	\rho^* = 
\bep
	\rhoZZ& \rhoZO\\
	\rhoOZ & \rhoOO\\
\eep
\eeq
is the density matrix of the qubit subspace. Here,  $\rho^*_x = \mathrm{Tr}[\rho^*\sigma_x]= 2\reRhoZO$, $\rho^*_y = \mathrm{Tr}[\rho^*\sigma_y]= -2\imRhoZO$, and $\rho^*_z = \mathrm{Tr}[\rho^*\sigma_z]= \rhoZZ - \rhoOO$.


\begin{thebibliography}{10}
\expandafter\ifx\csname url\endcsname\relax
  \def\url#1{\texttt{#1}}\fi
\expandafter\ifx\csname urlprefix\endcsname\relax\def\urlprefix{URL }\fi
\providecommand{\bibinfo}[2]{#2}
\providecommand{\eprint}[2][]{\url{#2}}

\bibitem{nakamura1999coherent}
\bibinfo{author}{Nakamura, Y.}, \bibinfo{author}{Pashkin, Y.} \&
  \bibinfo{author}{Tsai, J.}
\newblock \bibinfo{title}{Coherent control of macroscopic quantum states in a
  single-cooper-pair box}.
\newblock \emph{\bibinfo{journal}{Nature}} \textbf{\bibinfo{volume}{398}},
  \bibinfo{pages}{786--788} (\bibinfo{year}{1999}).

\bibitem{Chiorescu21032003}
\bibinfo{author}{Chiorescu, I.}, \bibinfo{author}{Nakamura, Y.},
  \bibinfo{author}{Harmans, C.~J.~P.~M.} \& \bibinfo{author}{Mooij, J.~E.}
\newblock \bibinfo{title}{Coherent quantum dynamics of a superconducting flux
  qubit}.
\newblock \emph{\bibinfo{journal}{Science}} \textbf{\bibinfo{volume}{299}},
  \bibinfo{pages}{1869--1871} (\bibinfo{year}{2003}).

\bibitem{PhysRevLett.89.117901}
\bibinfo{author}{Martinis, J.~M.}, \bibinfo{author}{Nam, S.},
  \bibinfo{author}{Aumentado, J.} \& \bibinfo{author}{Urbina, C.}
\newblock \bibinfo{title}{Rabi oscillations in a large josephson-junction
  qubit}.
\newblock \emph{\bibinfo{journal}{Phys.~Rev.~Lett.}}
  \textbf{\bibinfo{volume}{89}}, \bibinfo{pages}{117901}
  (\bibinfo{year}{2002}).

\bibitem{PhysRevA.76.042319}
\bibinfo{author}{Koch, J.} \emph{et~al.}
\newblock \bibinfo{title}{Charge-insensitive qubit design derived from the
  cooper pair box}.
\newblock \emph{\bibinfo{journal}{Phys.~Rev.~A}} \textbf{\bibinfo{volume}{76}},
  \bibinfo{pages}{042319} (\bibinfo{year}{2007}).
  
 \bibitem{2012arXiv1211.0322}
\bibinfo{author}{{Merkel}, S.~T.} \emph{et~al.}
\newblock \bibinfo{title}{{Self-consistent quantum process tomography}}.
\newblock {\bibinfo{journal}{Preprint at $<$http://arXiv.org/abs/1211.0322$>$}}
  (\bibinfo{year}{2012}).

\bibitem{PettaScience}
\bibinfo{author}{Petta, J.~R.} \emph{et~al.}
\newblock \bibinfo{title}{Coherent manipulation of coupled electron spins in
  semiconductor quantum dots}.
\newblock \emph{\bibinfo{journal}{Science}} \textbf{\bibinfo{volume}{309}},
  \bibinfo{pages}{2180--2184} (\bibinfo{year}{2005}).

\bibitem{koppens2006driven}
\bibinfo{author}{Koppens, F.~H.~L.} \emph{et~al.}
\newblock \bibinfo{title}{Driven coherent oscillations of a single electron
  spin in a quantum dot}.
\newblock \emph{\bibinfo{journal}{Nature}} \textbf{\bibinfo{volume}{442}},
  \bibinfo{pages}{766--771} (\bibinfo{year}{2006}).

\bibitem{nowakSO}
\bibinfo{author}{Nowack, K.~C.}, \bibinfo{author}{Koppens, F.~H.~L.},
  \bibinfo{author}{Nazarov, Y.~V.} \& \bibinfo{author}{Vandersypen, L.~M.~K.}
\newblock \bibinfo{title}{Coherent control of a single electron spin with
  electric fields}.
\newblock \emph{\bibinfo{journal}{Science}} \textbf{\bibinfo{volume}{318}},
  \bibinfo{pages}{1430--1433} (\bibinfo{year}{2007}).

\bibitem{gaudreau2011coherent}
\bibinfo{author}{Gaudreau, L.} \emph{et~al.}
\newblock \bibinfo{title}{Coherent control of three-spin states in a triple
  quantum dot}.
\newblock \emph{\bibinfo{journal}{Nature Phys.}} \textbf{\bibinfo{volume}{8}},
  \bibinfo{pages}{54--58} (\bibinfo{year}{2011}).

\bibitem{nadj2010spin}
\bibinfo{author}{Nadj-Perge, S.}, \bibinfo{author}{Frolov, S.},
  \bibinfo{author}{Bakkers, E.} \& \bibinfo{author}{Kouwenhoven, L.~P.}
\newblock \bibinfo{title}{Spin-orbit qubit in a semiconductor nanowire}.
\newblock \emph{\bibinfo{journal}{Nature}} \textbf{\bibinfo{volume}{468}},
  \bibinfo{pages}{1084--1087} (\bibinfo{year}{2010}).

\bibitem{LairdESR}
\bibinfo{author}{Laird, E.~A.} \emph{et~al.}
\newblock \bibinfo{title}{Hyperfine-mediated gate-driven electron spin
  resonance}.
\newblock \emph{\bibinfo{journal}{Phys.~Rev.~Lett.}}
  \textbf{\bibinfo{volume}{99}}, \bibinfo{pages}{246601}
  (\bibinfo{year}{2007}).

\bibitem{pioro2008electrically}
\bibinfo{author}{Pioro-Ladriere, M.} \emph{et~al.}
\newblock \bibinfo{title}{Electrically driven single-electron spin resonance in
  a slanting zeeman field}.
\newblock \emph{\bibinfo{journal}{Nature~Phys.}} \textbf{\bibinfo{volume}{4}},
  \bibinfo{pages}{776--779} (\bibinfo{year}{2008}).

\bibitem{folettiTomography}
\bibinfo{author}{Foletti, S.}, \bibinfo{author}{Bluhm, H.},
  \bibinfo{author}{Mahalu, D.}, \bibinfo{author}{Umansky, V.} \&
  \bibinfo{author}{Yacoby, A.}
\newblock \bibinfo{title}{Universal quantum control of two-electron spin
  quantum bits using dynamic nuclear polarization}.
\newblock \emph{\bibinfo{journal}{Nature Phys.}} \textbf{\bibinfo{volume}{5}},
  \bibinfo{pages}{903--908} (\bibinfo{year}{2009}).

\bibitem{pettaZener}
\bibinfo{author}{Petta, J.~R.}, \bibinfo{author}{Lu, H.} \&
  \bibinfo{author}{Gossard, A.~C.}
\newblock \bibinfo{title}{A coherent beam splitter for electronic spin states}.
\newblock \emph{\bibinfo{journal}{Science}} \textbf{\bibinfo{volume}{327}},
  \bibinfo{pages}{669--672} (\bibinfo{year}{2010}).

\bibitem{divincenzo2000universal}
\bibinfo{author}{DiVincenzo, D.~P.}, \bibinfo{author}{Bacon, D.},
  \bibinfo{author}{Kempe, J.}, \bibinfo{author}{Whaley, K.} \&
  \bibinfo{author}{Burkard, G.}
\newblock \bibinfo{title}{Universal quantum computation with the exchange
  interaction}.
\newblock \emph{\bibinfo{journal}{Nature}} \textbf{\bibinfo{volume}{408}},
  \bibinfo{pages}{339--342} (\bibinfo{year}{2000}).

\bibitem{LairdTdot}
\bibinfo{author}{Laird, E.~A.} \emph{et~al.}
\newblock \bibinfo{title}{Coherent spin manipulation in an exchange-only
  qubit}.
\newblock \emph{\bibinfo{journal}{Phys.~Rev.~B}} \textbf{\bibinfo{volume}{82}},
  \bibinfo{pages}{075403} (\bibinfo{year}{2010}).

\bibitem{hsieh2012physics}
\bibinfo{author}{Hsieh, C.}, \bibinfo{author}{Shim, Y.},
  \bibinfo{author}{Korkusinski, M.} \& \bibinfo{author}{Hawrylak, P.}
\newblock \bibinfo{title}{Physics of lateral triple quantum-dot molecules with
  controlled electron numbers}.
\newblock \emph{\bibinfo{journal}{Rep.~Prog.~Phys.}}
  \textbf{\bibinfo{volume}{75}}, \bibinfo{pages}{114501}
  (\bibinfo{year}{2012}).

\bibitem{2012arXiv1211.0417M}
\bibinfo{author}{{Mehl}, S.} \& \bibinfo{author}{{DiVincenzo}, D.~P.}
\newblock \bibinfo{title}{{Noise analysis of qubits implemented in triple
  quantum dot systems in a davies master equation approach}}.
\newblock {\bibinfo{journal}{Preprint at $<$http://arXiv.org/abs/1211.0417$>$}}
  (\bibinfo{year}{2012}).

\bibitem{west2012exchange}
\bibinfo{author}{West, J.~R.} \& \bibinfo{author}{Fong, B.~H.}
\newblock \bibinfo{title}{Exchange-only dynamical decoupling in the 3-qubit
  decoherence free subsystem}.
\newblock {\bibinfo{journal}{Preprint at $<$http://arXiv.org/abs/1203.4296$>$}}
  (\bibinfo{year}{2012}).

\bibitem{lundeen2008tomography}
\bibinfo{author}{Lundeen, J.} \emph{et~al.}
\newblock \bibinfo{title}{Tomography of quantum detectors}.
\newblock \emph{\bibinfo{journal}{Nature~Phys.}} \textbf{\bibinfo{volume}{5}},
  \bibinfo{pages}{27--30} (\bibinfo{year}{2008}).

\bibitem{brida2012quantum}
\bibinfo{author}{Brida, G.} \emph{et~al.}
\newblock \bibinfo{title}{Quantum characterization of superconducting photon
  counters}.
\newblock \emph{\bibinfo{journal}{New~J.~Phys.}}
  \textbf{\bibinfo{volume}{14}}, \bibinfo{pages}{085001}
  (\bibinfo{year}{2012}).

\bibitem{Buchachenko}
\bibinfo{author}{Buchachenko, A.~L.} \& \bibinfo{author}{Berdinsky, V.~L.}
\newblock \bibinfo{title}{Electron spin catalysis}.
\newblock \emph{\bibinfo{journal}{Chem.~Rev.}}
  \textbf{\bibinfo{volume}{102}}, \bibinfo{pages}{603--612}
  (\bibinfo{year}{2002}).

\bibitem{singletLikeFootnote}
\bibinfo{note}{The terms singlet-like and triplet-like refer to the spin state of two of the three electrons. For instance, for the singlet-like state $\SL$, the left two electrons are in a singlet state.}

\bibitem{ZeemanFootnote}
\bibinfo{note}{At the external magnetic field and electron temperature in this
  work, the $m_S=-1/2$ states are also loaded at times. These states behave
  identically to the $m_S=+1/2$ states in the regime presented in this work and
  states of different spin projections do not interact with one another. We
  have chosen therefore to ignore this added degree of freedom in the remainder
  of this work to reduce confusion.}

\bibitem{Reilly_APL07}
\bibinfo{author}{Reilly, D.~J.}, \bibinfo{author}{Marcus, C.~M.},
  \bibinfo{author}{Hanson, M.~P.} \& \bibinfo{author}{Gossard, A.~C.}
\newblock \bibinfo{title}{Fast single-charge sensing with a rf quantum point
  contact}.
\newblock \emph{\bibinfo{journal}{Appl.~Phys.~Lett.}}
  \textbf{\bibinfo{volume}{91}}, \bibinfo{pages}{162101}
  (\bibinfo{year}{2007}).

\bibitem{detuningFootnote}
\bibinfo{note}{$(\Vl^0,\Vm^0,\Vr^0) = (-727.09,-449.00,-301.86)$ mV.}

\bibitem{Cywinski_PRB08}
\bibinfo{author}{Cywi{\'n}ski, {\L}.}, \bibinfo{author}{Lutchyn, R.~M.},
  \bibinfo{author}{Nave, C.~P.} \& \bibinfo{author}{{Das Sarma}, S.}
\newblock \bibinfo{title}{How to enhance dephasing time in superconducting
  qubits}.
\newblock \emph{\bibinfo{journal}{Phys.~Rev.~B}} \textbf{\bibinfo{volume}{77}},
  \bibinfo{pages}{174509} (\bibinfo{year}{2008}).

\bibitem{AdiabatFootnote}
\bibinfo{note}{The values 5/8 and 3/8 only hold for pulses that are adiabatic
  with respect to the interdot tunnel couplings, but diabatic with respect to
  $\JL(0)+\JR(0)$, the total exchange in the center of 111. This allows pulses
  to be modeled as instantaneous changes of the Hamiltonian, giving $P_1 =
  |\bra{S_l}e^{-i\HJ{S}\tauJM{S}/\hbar}\ket{S_l}|^2 = 5/8 + 3/8
  \cos(\tauJM{S}\JR(\eps{S}{})/\hbar)$.}

\bibitem{Ladd12}
\bibinfo{author}{Ladd, T.~D.}
\newblock \bibinfo{title}{Hyperfine-induced decay in triple quantum dots}.
\newblock \emph{\bibinfo{journal}{Phys.~Rev.~B}} \textbf{\bibinfo{volume}{86}},
  \bibinfo{pages}{125408} (\bibinfo{year}{2012}).

\bibitem{ReillySpectrum}
\bibinfo{author}{Reilly, D.~J.} \emph{et~al.}
\newblock \bibinfo{title}{Measurement of temporal correlations of the
  overhauser field in a double quantum dot}.
\newblock \emph{\bibinfo{journal}{Phys.~Rev.~Lett.}}
  \textbf{\bibinfo{volume}{101}}, \bibinfo{pages}{236803}
  (\bibinfo{year}{2008}).

\bibitem{koppensPhaseShift}
\bibinfo{author}{Koppens, F.~H.~L.} \emph{et~al.}
\newblock \bibinfo{title}{Universal phase shift and nonexponential decay of
  driven single-spin oscillations}.
\newblock \emph{\bibinfo{journal}{Phys.~Rev.~Lett.}}
  \textbf{\bibinfo{volume}{99}}, \bibinfo{pages}{106803}
  (\bibinfo{year}{2007}).

\bibitem{BluhmT2}
\bibinfo{author}{Bluhm, H.} \emph{et~al.}
\newblock \bibinfo{title}{Dephasing time of gaas electron-spin qubits coupled
  to a nuclear bath exceeding 200 $\mu$s}.
\newblock \emph{\bibinfo{journal}{Nature~Phys.}} \textbf{\bibinfo{volume}{7}},
  \bibinfo{pages}{109--113} (\bibinfo{year}{2010}).

\bibitem{MedfordT2}
\bibinfo{author}{Medford, J.} \emph{et~al.}
\newblock \bibinfo{title}{Scaling of dynamical decoupling for spin qubits}.
\newblock \emph{\bibinfo{journal}{Phys.~Rev.~Lett.}}
  \textbf{\bibinfo{volume}{108}}, \bibinfo{pages}{086802}
  (\bibinfo{year}{2012}).

\bibitem{BarthelSingleShot}
\bibinfo{author}{Barthel, C.}, \bibinfo{author}{Reilly, D.~J.},
  \bibinfo{author}{Marcus, C.~M.}, \bibinfo{author}{Hanson, M.~P,} \&
  \bibinfo{author}{Gossard, A.~C.}
\newblock \bibinfo{title}{Rapid single-shot measurement of a singlet-triplet
  qubit}.
\newblock \emph{\bibinfo{journal}{Phys.~Rev.~Lett.}}
  \textbf{\bibinfo{volume}{103}}, \bibinfo{pages}{160503}
  (\bibinfo{year}{2009}).

\bibitem{POVMfootnote}
\bibinfo{note}{The four unknowns of $\ES{i}$ containing the qubit state-leakage
  state coherence are set to zero because without a coherent, stable magnetic
  field gradient, they are impossible to reliably measure and are negligibly
  small.}

\bibitem{POVMnoiseFootnote}
\bibinfo{note}{Noise on $\JL$ and $\JR$ was neglected, as $\eps{S}{3}$ and
  $\eps{S}{4}$ were in a region where exchange noise was not the dominant
  source of dephasing.}

\end{thebibliography}

\begin{thebibliography}{1}
\expandafter\ifx\csname url\endcsname\relax
  \def\url#1{\texttt{#1}}\fi
\expandafter\ifx\csname urlprefix\endcsname\relax\def\urlprefix{URL }\fi
\providecommand{\bibinfo}[2]{#2}
\providecommand{\eprint}[2][]{\url{#2}}

\bibitem{Reilly_APL07}
\bibinfo{author}{Reilly, D.~J.}, \bibinfo{author}{Marcus, C.~M.},
  \bibinfo{author}{Hanson, M.~P.} \& \bibinfo{author}{Gossard, A.~C.}
\newblock \bibinfo{title}{Fast single-charge sensing with a rf quantum point
  contact}.
\newblock \emph{\bibinfo{journal}{Appl.~Phys.~Lett.}}
  \textbf{\bibinfo{volume}{91}}, \bibinfo{pages}{162101}
  (\bibinfo{year}{2007}).

\bibitem{BarthelSingleShot}
\bibinfo{author}{Barthel, C.}, \bibinfo{author}{Reilly, D.~J.},
  \bibinfo{author}{Marcus, C.~M.}, \bibinfo{author}{Hanson, M.~P.} \&
  \bibinfo{author}{Gossard, A.~C.}
\newblock \bibinfo{title}{Rapid single-shot measurement of a singlet-triplet
  qubit}.
\newblock \emph{\bibinfo{journal}{Phys.~Rev.~Lett.}}
  \textbf{\bibinfo{volume}{103}}, \bibinfo{pages}{160503}
  (\bibinfo{year}{2009}).

\bibitem{Ladd12}
\bibinfo{author}{Ladd, T.~D.}
\newblock \bibinfo{title}{Hyperfine-induced decay in triple quantum dots}.
\newblock \emph{\bibinfo{journal}{Phys.~Rev.~B}} \textbf{\bibinfo{volume}{86}},
  \bibinfo{pages}{125408} (\bibinfo{year}{2012}).

\bibitem{LairdTdot}
\bibinfo{author}{Laird, E.~A.} \emph{et~al.}
\newblock \bibinfo{title}{Coherent spin manipulation in an exchange-only
  qubit}.
\newblock \emph{\bibinfo{journal}{Phys.~Rev.~B}} \textbf{\bibinfo{volume}{82}},
  \bibinfo{pages}{075403} (\bibinfo{year}{2010}).

\bibitem{SUPPmixedfootnote}
\bibinfo{note}{The $\Q$ mixing is confirmed by allowing for preparation
  infidelity in the MLE routine, and finding that the most likely population of
  the leakage state is 0.336, as opposed to the ideal 0.333. The ideal value is
  used in the data presented.}

\bibitem{PhysRevLett.108.070502}
\bibinfo{author}{Smolin, J.~A.}, \bibinfo{author}{Gambetta, J.~M.} \&
  \bibinfo{author}{Smith, G.}
\newblock \bibinfo{title}{Efficient method for computing the maximum-likelihood
  quantum state from measurements with additive gaussian noise}.
\newblock \emph{\bibinfo{journal}{Phys.~Rev.~Lett.}}
  \textbf{\bibinfo{volume}{108}}, \bibinfo{pages}{070502}
  (\bibinfo{year}{2012}).

\end{thebibliography}
\end{document}